\documentclass[a4paper,onecolumn,11pt,accepted=2024-04-07]{quantumarticle}
\pdfoutput=1
\usepackage[utf8]{inputenc}
\usepackage[english]{babel}
\usepackage[T1]{fontenc}

\usepackage[backend=biber,citestyle=numeric-comp,bibstyle=numeric,sorting=none,sortcites=true,maxnames=5,minnames=4]{biblatex}
\usepackage[breaklinks]{hyperref}
\usepackage{amsmath,amssymb,amsfonts,physics,bbm}

\usepackage[boxed]{algorithm}
\usepackage[noend]{algorithmic}

\newtheorem{Theorem}{Theorem}
\newtheorem{Corollary}{Corollary}
\newtheorem{Lemma}{Lemma} 
\newtheorem{Def}{Definition}

\newcommand{\one}{\mathbbm{1}}
\newcommand{\SP}{\textup{SP}}
\def\R{\mathbb{R}}
\def\Z{\mathbb{Z}}
\newcommand{\Herm}{\text{Herm}}
\newcommand{\Cl}{\mathcal{C}\ell}
\newcommand{\Span}[1]{\ensuremath{ \langle #1 \rangle }}
\newcommand{\set}[1]{\ensuremath{ \lbrace #1 \rbrace }}
\newcommand{\Sp}{\text{Sp}}

\begin{document}
	\title{Hidden variable model for quantum computation with magic states on qudits of any dimension}
	
	\author{Michael Zurel}
	\affiliation{Department of Physics and Astronomy, University of British Columbia, Vancouver, Canada}
	\affiliation{Stewart Blusson Quantum Matter Institute, University of British Columbia, Vancouver, Canada}
	\orcid{0000-0003-0333-1174}
	
	\author{Cihan Okay}
	\affiliation{Department of Mathematics, Bilkent University, Ankara, Turkey}
	\orcid{0000-0001-8097-5227}
	
	\author{Robert Raussendorf}
	\affiliation{Stewart Blusson Quantum Matter Institute, University of British Columbia, Vancouver, Canada}
	\affiliation{Institute of Theoretical Physics, Leibniz University Hannover, Hannover, Germany}
	\orcid{0000-0003-4983-9213}
	
	\author{Arne Heimendahl}
	\affiliation{Department of Mathematics and Computer Science, University of Cologne, Cologne, Germany}
	\orcid{0000-0002-8366-953X}
	
	\date{}
	
	\maketitle
	
	\begin{abstract}
		It was recently shown that a hidden variable model can be constructed for universal quantum computation with magic states on qubits.  Here we show that this result can be extended, and a hidden variable model can be defined for quantum computation with magic states on qudits with any Hilbert space dimension.  This model leads to a classical simulation algorithm for universal quantum computation.
	\end{abstract}

	\section{Introduction}
	
	The field of quantum computation has seen an explosion of interest in recent years.  It is widely believed that the era of quantum advantage is upon us and that we are entering the realm of so-called Noisy Intermediate-Scale Quantum (NISQ) computation.  This view is evidenced by the impressive performance of quantum devices in recent hardware demonstrations~\cite{Google2019,Google2020a,Google2020b,Google2021,IBM2020,DWave2018,IonQ2019,IonQ2020a,IonQ2020b,IonQ2020c}.
	
	But in spite of the age of the field and the recent surge in interest, a key question at the heart of quantum computation remains without an entirely satisfying answer: what is the essential quantum resource that provides the speedup of quantum computation over classical computation?  This is clearly an important question as its resolution could inform the development of quantum hardware and the design of quantum computer architectures.
	
	One inroad to approaching this question comes from quantum computation with magic states (QCM)~\cite{GottesmanPreskill2001,BravyiKitaev2005}.  QCM---a universal model of quantum computation closely related to the circuit model---is one of the leading candidates for scalable fault-tolerant quantum computation~\cite{CampbellVuillot2017}.  In QCM, the allowed operations are restricted to a subset of unitary gates forming the Clifford group, as well as arbitrary Pauli measurements.  These operations by themselves are not universal for quantum computation.  In fact, any quantum circuit consisting of only these operations can be simulated efficiently on a classical computer~\cite{Gottesman1998,AaronsonGottesman2004}, and so with these operations alone no quantum computational speedup is possible.  Universality is restored in QCM through the inclusion of additional nonstabilizer quantum states to the input of the circuit.  Therefore, this model allows us to refine the question posed above.  Instead of asking broadly ``which nonclassical resources are required for a quantum computational speedup?'', we can focus on the quantum states and ask ``which states could provide a speedup in QCM?''
	
	A partial answer to this question is provided by the study of quasiprobability representations like the Wigner function~\cite{Wigner1932}.  The Wigner function is the closest quantum mechanical counterpart to the classical notion of a probability distribution over a phase space, but unlike a probability distribution it can take negative values making it a \emph{quasi}probability function.  Accordingly, negativity in the Wigner function has traditionally been considered an indicator distinguishing classically behaving quantum states from those that exhibit genuinely quantum features~\cite{Hudson1974,KenfackZyczkowski2004}.  When adapted to finite-dimensional quantum mechanics, the setting relevant for quantum computation, quantum states are represented by a discrete Wigner function~\cite{Wootters1987,GibbonsWootters2004,Galvao2005,CormickGalvaoGottesman2006,Gross2006,Gross20062,GrossPhD2008}---a quasiprobability function over a finite set (a generalized phase space) usually satisfying certain constraints~\cite{Stratonovich1956,BrifMann1998,WignerCC}.
	
	Veitch et al.~\cite{VeitchEmerson2012} showed that a necessary condition for a quantum computational speedup in QCM on odd-prime-dimensional qudits---quantum systems with odd prime Hilbert space dimension---is that the discrete Wigner function of the input state of the quantum circuit must take negative values (this result is easily extended to QCM on qudits with any odd dimension~\cite{Zurel2020}).  In particular, the amount of negativity in the discrete Wigner function quantifies the cost of classical simulation of a quantum computation~\cite{PashayanBartlett2015} with simulation being efficient if the Wigner function is nonnegative everywhere.  Since nonnegativity of the discrete Wigner function also implies the existence of a classical (noncontextual) hidden variable model (HVM) describing the computation~\cite{Spekkens2008,DelfosseRaussendorf2017}, this proves that two traditional notions of nonclassicality for quantum systems---Wigner negativity and failure of a classical HVM description---herald a quantum computational advantage over classical computation.  This aligns with work that shows contextuality is required for quantum advantage in other settings~\cite{Raussendorf2013,Howard2014,Shahandeh2021}.
	
	The usual discrete Wigner function cannot be used to extend this result to QCM on even-dimensional qudits, including arguably the most important case---QCM on qubits~\cite{Zhu2016,KaranjaiBartlett2018,WignerCC,SchmidPusey2021}.  That said, similar necessary conditions for quantum advantage in QCM have been proven based on other quasiprobability representations~\cite{DelfosseRaussendorf2015,BermejoVegaRaussendorf2017,RaussendorfBermejoVega2017,HowardCampbell2017,KociaLove2017,RaussendorfZurel2020}.  In all cases, negativity is required in the representation of states or measurements in order to describe universal quantum computation.
	
	Recently, a hidden variable model was defined which bucks this trend by representing all quantum states, operations, and measurements relevant for QCM on qubits using only classical (nonnegative) probabilities~\cite{ZurelRaussendorf2020}.  This model is structurally similar to previous quasiprobability representations (modulo absence of negativity) and leads to a classical simulation method for universal quantum computation based on sampling from the defining probability distributions. In this paper we show that this result can be significantly extended in that a nonnegative hidden variable model can be constructed for quantum computation with magic states on qudits of any dimension.  We also show that many of the properties of the qubit HVM also apply in the qudit case, for example, this model leads to a classical simulation algorithm for quantum computation, and it subsumes previously defined quasiprobability representations. Note that, although this model can simulate any quantum computation, the simulation is presumably inefficient in general.
	
	\medskip
	
	The remainder of this paper is structured as follows.  In Section~\ref{Section:Background} we review some background material on quantum computation with magic states and the stabilizer formalism for qudits.  In Section~\ref{Section:HVM} we define the hidden variable model alluded to above.  In Section~\ref{Section:SimAlg} we present a classical simulation algorithm for quantum computation with magic states based on sampling from the probability distributions that define the model.  In Section~\ref{Section:CNCVertices} we characterize a subset of the hidden variables of the model.  Finally, in Section~\ref{Section:PhiMap} we consider some properties of the qubit HVM~\cite{OkayRaussendorf2021} which extend to the qudit case.  We conclude with a discussion of the significance of these results in Section~\ref{Section:Discussion}.

	\section{Preliminaries}\label{Section:Background}
	
	The setting of this paper is quantum computation with magic states (QCM)~\cite{GottesmanPreskill2001,BravyiKitaev2005} on multiple qudits, i.e., on multiple $d$-level quantum systems.  This is a universal model of quantum computation in which computation proceeds through the application a sequence of Clifford gates and Pauli measurements on an initially prepared ``magic state''.  Formally, for a system of $n$ $d$-dimensional qudits, the allowed measurements are associated with elements of the generalized Pauli group, $P=\langle\omega,X_k,Z_k\;|\;1\le k\le n\rangle$ where $\omega=\exp(2\pi i/d)$ is a primitive $d^{th}$ root of unity and the local Pauli operators are the $d$-dimensional generalization of the standard qubit Pauli operators~\cite{Gottesman1999oddDClifford}, given by
	\begin{equation}
		X=\sum\limits_{j\in\mathbb{Z}_d}\ket{j+1}\bra{j}\quad\text{and}\quad Z=\sum\limits_{j\in\mathbb{Z}_d}\omega^j\ket{j}\bra{j}.
	\end{equation}
	Here $\mathbb{Z}_d$ is the ring of integers modulo $d$.  With overall phases modded out we have $\mathcal{P}:=P/\mathcal{Z}(P)\cong\mathbb{Z}_d^{2n}$ and we can parametrize the Pauli operators by $a=(a_Z,a_X)\in\mathbb{Z}_d^n\times\mathbb{Z}_d^n=:E$ as
	\begin{equation}\label{eq:PauliOperators}
		T_a=\mu^{\phi(a)}Z(a_Z)X(a_X)
	\end{equation}
	with $Z(a_Z)=\bigotimes_{k=1}^nZ^{a_Z[k]}$, $X(a_X)=\bigotimes_{k=1}^nX^{a_X[k]}$.  Here $\mu=\omega$ when $d$ is odd, and $\mu=\sqrt{\omega}$ when $d$ is even.  The phase function $\phi$ can be chosen freely subject to the constraint $(T_a)^d=\one$ for all $a\in E$.  This constraint forces the eigenvalues of the operators to be in $\{\omega^j\;|\;j\in\mathbb{Z}_d\}$.  For concreteness we can choose the phase function to be
	\begin{equation}\label{eq:PhaseConvention}
		\phi(a)=\begin{cases}
			-\langle a_Z|a_X\rangle\cdot2^{-1}\quad&\text{if $d$ is odd}\\
			-\langle a_Z|a_X\rangle\quad&\text{if $d$ is even}
		\end{cases}
	\end{equation}
	where the inner product $ \langle a_Z|a_X \rangle := \sum_{k = 1}^n a_Z[k]a_X[k] $ is computed mod $d$ if $d$ is odd, and mod $2d$ if $d$ is even.
	
	The gates of the model are the Clifford gates, which are drawn from the normalizer of the Pauli group in the unitary group (up to overall phases): 
	\begin{equation}
		\Cl=\mathcal{N}(\mathcal{P})/U(1).
	\end{equation}
	The Clifford gates are not required for quantum computational universality of this model since they can always be propagated past the Pauli measurements, conjugating them into other Pauli measurements~\cite{DelfosseRaussendorf2015,BravyiSmolin2016}.  After they are propagated past the final Pauli measurements they can be dropped since they no longer affect the statistics of the measurements.  In the following we include the Clifford gates anyway for completeness.
	
	The last primitive required for QCM is the preparation of so-called ``magic'' input states.  It is these states which allow for the universality of QCM.  If the set of input states in QCM were restricted to only include $n$-qudit stabilizer states then the model would not be universal, and in fact any circuit of this type could be efficiently simulated on a classical computer.  This is the result of the Gottesman-Knill theorem~\cite{Gottesman1998,Gottesman1999oddDClifford}.  Magic states are any nonstabilizer states which allow for universality in the QCM model.
	
	\medskip
	
	See Appendix~\ref{Appendix:Stabilizer} for more background on Pauli measurements, Clifford gates, and the stabilizer formalism for qudits.
	
	\pagebreak
	
	Before proceeding we need to introduce some additional notation. The symplectic inner product $[\cdot,\cdot]:E\times E\longrightarrow\mathbb{Z}_d$ defined by
	\begin{equation}\label{eq:SymplecticProduct}
		[a,b]:=\langle a_Z|b_X\rangle-\langle a_X|b_Z\rangle
	\end{equation}
	tracks the commutator of the generalized Pauli operators in the sense
	\begin{equation}\label{eq:PauliCommutator}
		[T_a,T_b]:=T_aT_bT_a^{-1}T_b^{-1}=\omega^{[a,b]}\one.
	\end{equation}
	Because of this correspondence we will say that elements $a,b\in E$ commute when $[a,b]=0$.  The Pauli group with phases modded out forms a normal subgroup of the Clifford group such that $\Cl/\mathcal{P}\cong\Sp(E)$ is the group of symplectic transformations on $E$.  The Clifford group acts on the Pauli group by conjugation as
	\begin{equation}\label{eq:CliffordAction}
		gT_ag^\dagger=\omega^{\tilde{\Phi}_g(a)}T_{S_g(a)}\quad\forall g\in\Cl\;\forall a\in E
	\end{equation}
	where $S_g\in\Sp(E)$ is a symplectic map, and the function $\tilde{\Phi}_g:E\rightarrow\mathbb{Z}_d$ tracks the extra phases that get picked up.
	
	A function $\beta:E\times E\rightarrow\mathbb{Z}_d$ tracks how Pauli operators compose through the relation
	\begin{equation}
		T_aT_b=\omega^{-\beta(a,b)}T_{a+b}.
	\end{equation}
	The functions $\tilde{\Phi}$ and $\beta$ have a cohomological interpretation elucidated in Ref.~\cite{OkayRaussendorf2017} (also see Refs.~\cite{RaussendorfZurel2020,WignerCC}).
	
	\medskip
	
	Recall from Ref.~\cite{RaussendorfZurel2020} a few definitions.
	\begin{Def}
		A set $\Omega\subset E$ is \emph{closed under inference} if for every pair of elements $a,b\in\Omega$ satisfying $[a,b]=0$, it holds that $a+b\in\Omega$.  The \emph{closure under inference} of a set $\Omega\subset E$, denoted $\bar{\Omega}$, is the smallest subset of $E$ which contains $\Omega$ and is closed under inference.
	\end{Def}
	\begin{Def}
		A set $\Omega\subset E$ is called \emph{noncontextual} if there exists a noncontextual value assignment for the closure of that set, i.e., there exists a function $\gamma:\bar{\Omega}\rightarrow\mathbb{Z}_d$ such that for all $a,b\in\Omega$ with $[a,b]=0$, $\gamma$ satisfies
		\begin{equation}\label{eq:noncontextuality condition}
			\gamma(a)+\gamma(b)-\gamma(a+b)=-\beta(a,b).
		\end{equation}
	\end{Def}
	A set which is both closed under inference and noncontextual we call cnc for short.
	
	For any isotropic subgroup $I\subset E$, i.e.~any subgroup $I$ of $E$ on which the symplectic product is identically zero, and any noncontextual value assignment $r:I\rightarrow\mathbb{Z}_d$, the operator
	\begin{equation}
		\Pi_I^r=\frac{1}{|I|}\sum\limits_{b\in I}\omega^{-r(b)}T_b
	\end{equation}
	is the projector onto the simultaneous $+1$-eigenspace of the operators $\{\omega^{-r(b)}T_b\;|\;b\in I\}$.  This represents a measurement of the Pauli observables labeled by $I$ yielding measurement outcomes $\omega^{r(b)}$.  In particular, for a single Pauli measurement $I=\langle a\rangle$ is generated by a single element $a\in E$, and when $|I|=d^n$, $\Pi_I^r$ is a projector onto a stabilizer state.

	\section{Hidden variable model}\label{Section:HVM}
	
	In this section we define a hidden variable model that represents all components of quantum computation with magic states by a family of probability distributions.  This is in contrast to previous quasiprobability representations which required negativity in the representation of either the states or the operations of QCM in order to represent universal quantum computation.  The main result of this section is Theorem~\ref{Theorem:HVMDef}.
	
	\medskip
	
	Let $\Herm(\mathcal{H})$ be the space of Hermitian operators on $n$-qudit Hilbert space $\mathcal{H}\cong\mathbb{C}^{d^n}$, $\Herm_1(\mathcal{H})$ be the affine subspace of this space obtained by fixing the trace of the operators to be $1$, and let $\Herm_1^{\succeq0}(\mathcal{H})$ be the subset of $\Herm_1(\mathcal{H})$ consisting of positive semidefinite operators.  Let $\mathcal{S}$ denote the set of pure $n$-qudit stabilizer states.  The state space of the hidden variable model is based on the set
	\begin{equation}\label{eq:LambdaDef}
		\Lambda=\{X\in\Herm_1(\mathcal{H})|\Tr(\ket{\sigma}\bra{\sigma}X)\ge0\;\forall\ket{\sigma}\in\mathcal{S}\}.
	\end{equation}
	The elements of $\Lambda$ are much like density operators in that they are Hermitian operators with unit trace, but unlike density operators they are not necessarily positive semidefinite.  In order to define the hidden variable model we first need to establish some basic properties of $\Lambda$.
	
	\begin{Lemma}\label{lemma:compactness-of-Lambda}
		For any number of qudits $n\in\mathbb{N}$ of any dimension $d\in\mathbb{N}$, (i)~$\Lambda$ is convex, and (ii)~$\Lambda$ is compact.
	\end{Lemma}
	The proof of Lemma~\ref{lemma:compactness-of-Lambda} is in Appendix~\ref{Appendix:Lemma1}.
	
	\medskip
	
	$\Lambda$ can be interpreted as a subset of a real affine space defined by the intersection of a finite number of linear inequalities, i.e., it is a polyhedral set.  Since $\Lambda$ is compact, it is a polytope, and so by the Minkowski-Weyl theorem~\cite{Ziegler1995} it can equivalently be described as the convex hull of finitely many vertices.  Let $\{A_\alpha\;|\;\alpha\in\mathcal{V}\}$ denote the (finite) set of vertices of $\Lambda$.  Then we have the following result, which is a generalization of \cite[Theorem~1]{Zurel2020} to qudits of arbitrary Hilbert space dimension $ d $.
	\begin{Theorem}\label{Theorem:HVMDef}
		For any number $n\in\mathbb{N}$ of qudits with any Hilbert space dimension $d\in\mathbb{N}$,
		\begin{enumerate}
			\item{For any quantum state $\rho\in\Herm_1^{\succeq0}(\mathcal{H})$, there is a probability function $p_\rho:\mathcal{V}\rightarrow\mathbb{R}_{\ge0}$ such that
				\begin{equation}\label{eq:HVMStateExpansion}
					\rho=\sum\limits_{\alpha\in\mathcal{V}}p_\rho(\alpha)A_\alpha.
				\end{equation}
			}
			\item{For any vertex $A_\alpha$ of $\Lambda$ and any Clifford operation $g\in\Cl$, $gA_\alpha g^\dagger=:A_{g\cdot \alpha}$ is a vertex of $\Lambda$.
			}
			\item{For update under Pauli measurements it holds that for any isotropic subgroup $I\subset E$, any noncontextual value assignment $r:I\rightarrow\mathbb{Z}_d$, and any vertex $A_\alpha$,
				\begin{equation}\label{eq:HVMPauliUpdate}
					\Pi_I^rA_\alpha\Pi_I^r=\sum\limits_{\alpha'\in\mathcal{V}}q_{\alpha,I}(\alpha',r)A_{\alpha'},
				\end{equation}
				where $q_{\alpha,I}(\alpha',r)\ge0$ for all $\alpha'\in\mathcal{V}$ and $\sum_{\alpha',r}q_{\alpha,I}(\alpha',r)=1$.
			}
			\item{The Born rule takes the form
				\begin{equation}
					\Tr(\Pi_I^r\rho)=\sum\limits_{\alpha\in\mathcal{V}}p_\rho(\alpha)Q_I(r\;|\;\alpha)
				\end{equation}
				where $Q$ is given by
				\begin{equation}
					Q_I(r\;|\;\alpha)=\sum\limits_{\alpha'\in\mathcal{V}}q_{\alpha,I}(\alpha',r).
				\end{equation}
			}
		\end{enumerate}
	\end{Theorem}
	
	This theorem defines a hidden variable model which represents all of the primitives of quantum computation with magic states using only probability distributions for states and measurements, and probabilistic update rules for dynamics.  It has a similar structure to previous hidden variable models based on quasiprobability representations~\cite{Gross2006,Gross20062,GrossPhD2008,DelfosseRaussendorf2015,PashayanBartlett2015,DelfosseRaussendorf2017,HowardCampbell2017,DeBrotaStacey2020,RaussendorfZurel2020,ZurelRaussendorf2020,Zurel2020}, with a key difference being that in this model every state can be represented by a probability distribution, no negativity is required.  Unlike the model of Beltrametti and Bugajski~\cite{BeltramettiBugajski1996} which requires a hidden variable for each pure quantum state, this model has a finite number of hidden variables for any number of qudits.
	
	\medskip
	
	The proof of Theorem~\ref{Theorem:HVMDef} requires the following lemma.
	\begin{Lemma}\label{Lemma:polytopeProperties}
		The polytope $\Lambda$ has the following properties:
		\begin{enumerate}
			\item For any density operator $\rho\in\Herm_1^{\succeq0}(\mathcal{H})$ representing a physical $n$-qudit quantum state, $\rho\in\Lambda$.
			\item For any $X\in\Lambda$ and any Clifford operation $g\in\Cl$, $gXg^\dagger\in\Lambda$,
			\item For any $X\in\Lambda$, any isotropic subgroup $I\subset E$, and any noncontextual value assignment $r:I\rightarrow \mathbb{Z}_d$, if $\Tr(\Pi_I^rX)>0$ then 
			\begin{equation}
				\frac{\Pi_I^rX\Pi_I^r}{\Tr(\Pi_I^rX)}\in\Lambda.
			\end{equation}
		\end{enumerate}
	\end{Lemma}
	
	\emph{Proof of Lemma~\ref{Lemma:polytopeProperties}.} We will prove the three properties of the lemma in order.  For the first property note that for any density operator $\rho\in\Herm_1^{\succeq0}(\mathcal{H})$, $\rho$ is positive semidefinite.  Therefore, for any pure quantum state $\ket{\psi}$, $\Tr(\ket{\psi}\bra{\psi}\rho)\ge0$.  This holds in particular for any pure stabilizer state.  Therefore, $\rho$ satisfies all of the defining inequalities of the polytope in eq.~\eqref{eq:LambdaDef} and so $\rho\in\Lambda$.
	
	For the second property, let $X\in\Lambda$ and $g\in\Cl$.  Then for any stabilizer state $\ket{\sigma}\in S$,
	\begin{align}
		\Tr(\ket{\sigma}\bra{\sigma}(gXg^\dagger))=&\Tr((g^\dagger\ket{\sigma}\bra{\sigma}g)X)\\=&\Tr(\ket{\sigma'}\bra{\sigma'}X)\ge0.
	\end{align}
	Here the first equality follows from the cyclic property of the trace, the second equality from the fact that Clifford operations map stabilizer states to stabilizer states (see Lemma~\ref{Lemma:StabilizerClifford} in Appendix~\ref{Appendix:Stabilizer}), and the last inequality from the assumption $X\in\Lambda$.
	
	Now we can prove the third property of Lemma~\ref{Lemma:polytopeProperties}.  Let $I,J\subset E$ be isotropic subgroups with noncontextual value assignments $r:I\rightarrow\mathbb{Z}_d$ and $s:J\rightarrow\mathbb{Z}_d$.  By Lemma~\ref{Lemma:StabilizerPauli} in Appendix~\ref{Appendix:Stabilizer},
	\begin{equation}
		\Pi_I^r\Pi_J^s\Pi_I^r=\delta_{r|_{I\cap J},s|_{I\cap J}}\frac{|J\cap I^\perp|}{|J|}\Pi_{I+(J\cap I^\perp)}^{r\star s}
	\end{equation}
	where $\delta_{r|_{I\cap J},s|_{I\cap J}}$ is equal to one if $r$ and $s$ agree on the intersection $I\cap J$, and it is zero otherwise. Here $r\star s$ is the unique noncontextual value assignment on the set $I+(J\cap I^\perp)$ such that $r\star s|_I=r$ and $r\star s|_{J\cap I^\perp}=s|_{J\cap I^\perp}$.  For any $X\in\Lambda$ and any Pauli projector $\Pi_I^r$, if $\Tr(\Pi_I^rX)>0$, then for any projector onto a stabilizer state $\Pi_J^s$,
	\begin{align}
		\Tr(\Pi_J^s\frac{\Pi_I^rX\Pi_I^r}{\Tr(\Pi_I^rX)})&=\frac{\Tr((\Pi_I^r\Pi_J^s\Pi_I^r)X)}{\Tr(\Pi_I^rX)}\\
		&=\delta_{r|_{I\cap J},s|_{I\cap J}}\frac{|J\cap I^\perp|}{|J|}\frac{\Tr(\Pi_{I+(J\cap I^\perp)}^{r\star s}X)}{\Tr(\Pi_I^r X)}\ge0.
	\end{align}
	Here the first line follows from linearity and the cyclic property of the trace.   The last inequality follows from the fact that by Lemma~\ref{lemma:SumsOfStabilizer CodeProjectors} in Appendix~\ref{Appendix:Stabilizer}, $\Pi_{I+(J\cap I^\perp)}^{r\star s}$ can be written as a conic combination of projectors onto stabilizer states, and from the assumption $X\in\Lambda$.  Therefore, for any $X\in\Lambda$, any Pauli projector $\Pi_I^r$, and any stabilizer state $\ket{\sigma}\in\mathcal{S}$, if $\Tr(\Pi_I^rX)>0$, then
	\begin{equation}
		\Tr(\ket{\sigma}\bra{\sigma}\frac{\Pi_I^rX\Pi_I^r}{\Tr(\Pi_I^rX)})\ge0
	\end{equation}
	and so $\Pi_I^rX\Pi_I^r/\Tr(\Pi_I^rX)\in\Lambda$.  This proves the third statement of the lemma.
	$\Box$
	
	\medskip
	
	We can now prove the main result of this section.
	
	\emph{Proof of Theorem~\ref{Theorem:HVMDef}.} We will prove the four statements of the theorem in order.  First, as shown in Lemma~\ref{Lemma:polytopeProperties}, $\Lambda$ contains all density matrices corresponding to physical $n$-qudit quantum states.  Therefore, by the Krein-Milman theorem~\cite{Ziegler1995} any state can be written as a convex combination of the vertices of $\Lambda$.  This is the first statement of the theorem.
	
	The second property from Lemma~\ref{Lemma:polytopeProperties} shows that for any Clifford operation $g\in\Cl$ and any vertex $A_\alpha$ of $\Lambda$, we have $A_{g\cdot\alpha}:=gA_\alpha g^\dagger\in\Lambda$.  It remains to show that $A_{g\cdot\alpha}$ is a vertex of $\Lambda$.
	
	Let $\mathcal{S}_\alpha=\{\ket{\sigma}\in\mathcal{S}\;|\;\Tr(\ket{\sigma}\bra{\sigma}A_\alpha)=0\}$ be the set of stabilizer states with projectors orthogonal to vertex $A_\alpha$ with respect to the Hilbert-Schmidt inner product.  By Theorem~18.1 of Ref.~\cite{Chvatal1983}, since $A_\alpha$ is a vertex of $\Lambda$, $A_\alpha$ is the unique solution in $\Herm_1(\mathcal{H})$ of the system
	\begin{equation}\label{eq:vertexSOE}
		\begin{cases}
			\Tr(\ket{\sigma}\bra{\sigma}X)=0\quad\forall\ket{\sigma}\in\mathcal{S}_\alpha\\
			\Tr(\ket{\sigma}\bra{\sigma}X)\ge0\quad\forall\ket{\sigma}\in\mathcal{S}\setminus\mathcal{S}_\alpha.
		\end{cases}
	\end{equation}
	For any stabilizer state $\ket{\sigma}\in\mathcal{S}_\alpha$,
	\begin{align}
		\Tr(\ket{\sigma}\bra{\sigma}X)=&\Tr(g\ket{\sigma}\bra{\sigma}g^\dagger gXg^\dagger).
	\end{align}
	Therefore, under conjugation by $g\in\Cl$, solutions to the system eq.~\eqref{eq:vertexSOE} are mapped bijectively to solutions of the system
	\begin{equation}\label{eq:vertexSOEClifford}
		\begin{cases}
			\Tr(\ket{\sigma}\bra{\sigma}X)=0\quad\forall\ket{\sigma}\in\mathcal{S}_{g\cdot\alpha}\\
			\Tr(\ket{\sigma}\bra{\sigma}X)\ge0\quad\forall\ket{\sigma}\in\mathcal{S}\setminus\mathcal{S}_{g\cdot\alpha}
		\end{cases}
	\end{equation}
	where $S_{g\cdot\alpha}:=\{g\ket{\sigma}\;|\;\ket{\sigma}\in S_\alpha\}$.  In particular, $A_{g\cdot\alpha}:=gA_\alpha g^\dagger$ is the unique solution to this system, so by Theorem~18.1 of Ref.~\cite{Chvatal1983}, it is a vertex of $\Lambda$.
	
	For the third statement of Theorem~\ref{Theorem:HVMDef}, let $A_\alpha$ be a vertex of $\Lambda$ and $\Pi_I^r$ be any Pauli projector.  We have two cases.  (I)~First, if $\Tr(\Pi_I^rA_\alpha)=0$, then $\Pi_I^rA_\alpha\Pi_I^r$ is zero as an operator.  To see this, consider the inner product $\Tr(T_a\Pi_I^rA_\alpha\Pi_I^r)$ for any Pauli operator $T_a,\;a\in E$.  Here we have three subcases: (i) if $a\in I$, then
	\begin{align}
		T_a\Pi_I^r=&\frac{1}{|I|}\sum\limits_{b\in I}\omega^{-r(b)}T_aT_b\\
		=&\frac{1}{|I|}\sum\limits_{b\in I}\omega^{-r(b)-\beta(a,b)}T_{a+b}\\
		=&\frac{1}{|I|}\sum\limits_{b\in I}\omega^{r(a)-r(a+b)}T_{a+b}\\
		=&\omega^{r(a)}\Pi_I^r.
	\end{align}
	Therefore,
	\begin{align}
		\Tr(T_a\Pi_I^rA_\alpha\Pi_I^r)=\omega^{r(a)}\Tr(\Pi_I^rA_\alpha)=0.
	\end{align}
	(ii) If $a\in I^{\perp}\setminus I$, then by Lemma~\ref{lemma:SumsOfStabilizer CodeProjectors} in Appendix~\ref{Appendix:Stabilizer} we have
	\begin{equation}
		\Pi_I^r=\sum\limits_{\gamma\in\Gamma_{I,r}^{\langle a,I\rangle}}\Pi_{\langle a,I\rangle}^\gamma
	\end{equation}
	where $\Gamma_{I,r}^{\langle a,I\rangle}$ is the set of noncontextual value assignments on $\langle a,I\rangle$ satisfying $\gamma|_I=r$.  Multiplying this equation on the right by $A_\alpha$ and taking a trace we get
	\begin{align}
		\Tr(\Pi_I^rA_\alpha)=\sum\limits_{\gamma\in\Gamma_{I,r}^{\langle a,I\rangle}}\Tr(\Pi_{\langle a,I\rangle}^\gamma A_\alpha).
	\end{align}
	With Lemma~\ref{lemma:SumsOfStabilizer CodeProjectors}, each projector on the right hand side can be written as a sum of projectors onto stabilizer states.  Therefore, since $A_\alpha\in\Lambda$ each term on the right hand side is nonnegative.  But the left hand side is zero by assumption.  Therefore, each term on the right hand side is zero.  I.e. $\Tr(\Pi_{\langle a,I\rangle}^\gamma A_\alpha)=0$ for every $\gamma\in\Gamma_{I,r}^{\langle a,I\rangle}$.
	
	We can write the spectral decomposition of the operator $T_a$ as
	\begin{equation}
		T_a=\sum\limits_{\gamma\in\Gamma_{I,r}^{\langle a,I\rangle}}\omega^{\gamma(a)}\Pi_{\langle a\rangle}^{\gamma|_{\langle a\rangle}}.
	\end{equation}
	Then
	\begin{align}
		\Tr(T_a\Pi_I^rA_\alpha\Pi_I^r)=&\sum\limits_{\gamma\in\Gamma_{I,r}^{\langle a,I\rangle}}\omega^{\gamma(a)}\Tr(\Pi_{\langle a\rangle}^{\gamma|_{\langle a\rangle}}\Pi_I^rA_\alpha\Pi_I^r)\\
		=&\sum\limits_{\gamma\in\Gamma_{I,r}^{\langle a,I\rangle}}\omega^{\gamma(a)}\Tr(\Pi_{\langle a,I\rangle}^\gamma A_\alpha)=0.
	\end{align}
	
	(iii) If $a\not\in I^\perp$, then
	\begin{align}
		\Pi_I^rT_a\Pi_I^r&=\frac{1}{|I|^2}\sum\limits_{b,c\in I}\omega^{-r(b)-r(c)}T_bT_aT_c\\
		=&\frac{1}{|I|^2}\sum\limits_{b,c\in I}\omega^{-r(b)-r(c)+[b,a]-\beta(b,c)}T_aT_{b+c}\\
		=&\frac{1}{|I|^2}T_a\sum\limits_{b,c\in I}\omega^{-r(b+c)+[b,a]}T_{b+c}\\
		=&\frac{1}{|I|}T_a\Pi_I^r\sum\limits_{b\in I}\omega^{[b,a]}.
	\end{align}
	By character orthogonality, the sum in the final expression vanishes.  Therefore,
	\begin{equation}
		\Tr(T_a\Pi_I^rA_\alpha\Pi_I^r)=\Tr(\Pi_I^rT_a\Pi_I^rA_\alpha)=0.
	\end{equation}

	We have $\Tr(T_a\Pi_I^rA_\alpha\Pi_I^r)=0$ for every Pauli operator $T_a,\;a\in E$.  Therefore, $\Pi_I^rA_\alpha\Pi_I^r$ is zero as an operator.
	
	(II)~Second, if $\Tr(\Pi_I^rA_\alpha)>0$, then by the third statement of Lemma~\ref{Lemma:polytopeProperties} we have $\Pi_I^rA_\alpha\Pi_I^r/\Tr(\Pi_I^rA_\alpha)\in\Lambda$, and so there exists a decomposition of $\Pi_I^rA_\alpha\Pi_I^r/\Tr(\Pi_I^rA_\alpha)$ as a convex combination of the vertices of $\Lambda$.  Therefore, there exist nonnegative coefficients $q_{\alpha,I}(\alpha',r)$ such that
	\begin{equation}
		\Pi_I^rA_\alpha\Pi_I^r=\sum\limits_{\alpha'\in\mathcal{V}}q_{\alpha,I}(\alpha',r)A_{\alpha'}.
	\end{equation}
	Taking a trace of this equation and adding the corresponding equations for all noncontextual value assignments $r$ of $I$ we have on the left hand side
	\begin{equation}
		\sum\limits_{r}\Tr(\Pi_I^rA_\alpha\Pi_I^r)=\Tr\left[\left(\sum\limits_{r}\Pi_I^r\right)A_\alpha\right]=\Tr(A_\alpha)=1
	\end{equation}
	and on the right hand side
	\begin{equation}
		\sum_{\alpha',r}q_{\alpha,I}(\alpha',r)\Tr(A_{\alpha'})=\sum_{\alpha',r}q_{\alpha,I}(\alpha',r).
	\end{equation}
	Therefore, $\sum_{\alpha',r}q_{\alpha,I}(\alpha',r)=1$.  This proves the third statement of the Theorem.
	
	Finally, we calculate
	\begin{align}
		\Tr(\Pi_I^r\rho)=&\sum\limits_{\alpha\in\mathcal{V}}p_\rho(\alpha)\Tr(\Pi_I^rA_\alpha)\\
		=&\sum\limits_{\alpha\in\mathcal{V}}p_\rho(\alpha)\sum\limits_{\alpha'\in\mathcal{V}}q_{\alpha,I}(\alpha',r)\\
		=&\sum\limits_{\alpha\in\mathcal{V}}p_\rho(\alpha)Q_I(r\;|\;\alpha)
	\end{align}
	and we obtain the fourth statement of the theorem. $\Box$

	\section{Classical simulation algorithm}\label{Section:SimAlg}
	
	Theorem~\ref{Theorem:HVMDef} shows that all of the components of quantum computation with magic states can be described by a hidden variable model which represents all relevant states and dynamical operations by probabilities.  This leads to a classical simulation algorithm for quantum computation with magic states, Algorithm~\ref{Algorithm:simAlg}, based on sampling from these probability distributions.
	
	In short, a vertex label $\alpha\in\mathcal{V}$ is sampled according to the probability distribution of eq.~\eqref{eq:HVMStateExpansion} representing the input state of the quantum circuit.  This vertex is then propagated through the circuit.  When a Clifford gate $g\in\Cl$ is encountered, we have a deterministic update rule: $\alpha\rightarrow g\cdot\alpha$, according to the second statement of Theorem~\ref{Theorem:HVMDef}.  When a Pauli measurement $a\in E$ is encountered, the third and fourth statements of Theorem~\ref{Theorem:HVMDef} give a way of determining probabilities for measurement outcomes, as well as a probabilistic update rule.  That is, we sample a pair $(\alpha',r)$ according to the probability distribution $q_{\alpha,\langle a\rangle}$ where $\alpha'\in\mathcal{V}$ and $r:\langle a\rangle\rightarrow\mathbb{Z}_d$ is a noncontextual value assignment.  Then $r(a)$ is returned as the measurement outcome and the vertex is updated as $\alpha\rightarrow\alpha'$.  This process continues until the end of the circuit is reached.
	
	\begin{algorithm}[H]
		\begin{algorithmic}[1]
			\REQUIRE $p_{\rho_{0}}$
			\STATE sample a point $\alpha\in\mathcal{V}$ according to the probability distribution $p_{\rho_{0}}$
			\WHILE{end of circuit has not been reached}
			\IF{a Clifford gate $g\in\Cl$ is encountered}
			\STATE update $\alpha\leftarrow g\cdot\alpha$
			\ENDIF
			\IF{a Pauli measurement $T_a,\;a\in E$ is encountered}
			\STATE sample $(\alpha',r)$ according to the probability distribution $q_{\alpha,\langle a\rangle}$
			\STATE \textbf{Output:} $r(a)$ as the outcome of the measurement
			\STATE update $\alpha\leftarrow\alpha'$
			\ENDIF
			\ENDWHILE
		\end{algorithmic}
		\caption{One run of the classical simulation algorithm for quantum computation with magic states based on the hidden variable model of Theorem~\ref{Theorem:HVMDef}.  The algorithm provides samples from the joint probability distribution of the Pauli measurements in a quantum circuit consisting of Clifford unitaries and Pauli measurements applied to an input state $\rho_{0}$.\label{Algorithm:simAlg}}
	\end{algorithm}
	
	A proof of the correctness of this simulation algorithm is given below.
	
	\begin{Theorem}\label{Theorem:simAlgCorrect}
		The classical simulation algorithm, Algorithm~\ref{Algorithm:simAlg}, correctly reproduces the predictions of quantum theory.
	\end{Theorem}
	
	\emph{Proof of Theorem~\ref{Theorem:simAlgCorrect}.} Without loss of generality, a  QCM circuit can be represented as a sequence $g_1,I_1,g_2,I_2,\dots$ with $g_1,g_2,\dots\in\Cl$ specifying the Clifford unitaries to be applied, and $I_1,I_2,\dots\subset E$ isotropic subgroups specifying the Pauli measurements to be performed.  First, consider a single layer of this circuit consisting of a Clifford gate $g\in\Cl$ followed by Pauli measurements corresponding to an isotropic subgroup $I\subset E$.
	
	Using the classical simulation algorithm, Algorithm~\ref{Algorithm:simAlg}, the conditional probability of obtaining measurement outcomes specified by the noncontextual value assignment $r:I\rightarrow\mathbb{Z}_d$ for the measurements given the state $\alpha\in\mathcal{V}$ is $Q_I(r\;|\;g\cdot\alpha)$.  Therefore, the probability of obtaining outcomes $r$ given the gate $g$ is applied to the state $\rho$ followed by the measurements of the Pauli observables in $I$ is given by
	\begin{equation}
		P_{\rho,g,I}^{(Sim)}(r)=\sum\limits_{\alpha\in\mathcal{V}}p_\rho(\alpha)Q_I(r\;|\;g\cdot\alpha).
	\end{equation}
	The corresponding outcome probability predicted by the Born rule, $P_{\rho,g,I}^{(QM)}(r)$, is
	\begin{align}
		\Tr(\Pi_I^rg\rho g^\dagger)=&\sum\limits_{\alpha\in\mathcal{V}}p_\rho(\alpha)\Tr(\Pi_I^rgA_\alpha g^\dagger)\\
		=&\sum\limits_{\alpha\in\mathcal{V}}p_\rho(\alpha)\Tr(\Pi_I^rA_{g\cdot\alpha})\\
		=&\sum\limits_{\alpha\in\mathcal{V}}p_\rho(\alpha)Q_I(r\;|\;g\cdot\alpha).
	\end{align}
	Here in the first line we use the expansion of $\rho$ in the vertices of $\Lambda$, eq.~\eqref{eq:HVMStateExpansion}, in the second line we use the second statement of Theorem~\ref{Theorem:HVMDef}, and in the last line we use the fourth statement of Theorem~\ref{Theorem:HVMDef}.  This agrees with the outcome probability predicted by the classical simulation algorithm.
	
	Now consider the postmeasurement state $\rho'$.  According to quantum mechanics, the postmeasurement state is
	\begin{equation}
		\rho'^{(QM)}=\frac{\Pi_I^rg\rho g^\dagger\Pi_I^r}{\Tr(\Pi_I^rg\rho g^\dagger)}.
	\end{equation}
	Here the numerator is
	\begin{align}
		\Pi_I^rg\rho g^\dagger\Pi_I^r=&\sum\limits_{\alpha\in\mathcal{V}}p_\rho(\alpha)\Pi_I^rgA_\alpha g^\dagger\Pi_I^r\\
		=&\sum\limits_{\alpha\in\mathcal{V}}p_\rho(\alpha)\Pi_I^rA_{g\cdot\alpha}\Pi_I^r\\
		=&\sum\limits_{\alpha\in\mathcal{V}}p_\rho(\alpha)\sum\limits_{\alpha'\in\mathcal{V}}q_{g\cdot\alpha,I}(\alpha',r)A_{\alpha'}
	\end{align}
	and the denominator is
	\begin{align}
		\Tr(\Pi_I^rg\rho g^\dagger\Pi_I^r)=&\sum\limits_{\alpha\in\mathcal{V}}p_\rho(\alpha)\sum\limits_{\alpha'\in\mathcal{V}}q_{g\cdot\alpha,I}(\alpha',r)\\
		=&\sum\limits_{\alpha\in\mathcal{V}}p_\rho(\alpha)Q_I(r\;|\;g\cdot\alpha),
	\end{align}
	so the postmeasurement state predicted by quantum theory is
	\begin{equation}
		\rho'^{(QM)}=\frac{\sum_{\alpha}p_\rho(\alpha)\sum_{\alpha'}q_{g\cdot\alpha,I}(\alpha',r)A_{\alpha'}}{\sum_{\alpha}p_\rho(\alpha)Q_I(r\;|\;g\cdot\alpha)}.
	\end{equation}
	
	Using the classical simulation algorithm, the probability of obtaining outcomes $r$ and state $A_{\alpha'}$ given a Clifford gate $g$ followed by measurements of the Pauli observables $I$ on state $\rho$ is $P_{\rho,g,I}(\alpha',r)=P_{\rho,g,I}(\alpha'|r)P_{\rho,g,I}(r)$.  But $P_{\rho,g,I}(\alpha',r)=\sum_{\alpha}p_\rho(\alpha)P_{g,I}(\alpha',r|\alpha)=\sum_\alpha p_\rho(\alpha)q_{g\cdot\alpha,I}(\alpha',r)$, and $P_{\rho,g,I}(\alpha'|r)=p_{\rho'}(\alpha')$.  Therefore, the postmeasurement state predicted by the classical simulation algorithm is
	\begin{align}
		\rho'^{(Sim)}=&\sum\limits_{\alpha'\in\mathcal{V}}p_{\rho'}(\alpha')A_{\alpha'}\\
		=&\sum\limits_{\alpha'\in\mathcal{V}}\frac{P_{\rho,g,I}(\alpha',r)}{P_{\rho,g,I}(r)}A_{\alpha'}\\
		=&\sum\limits_{\alpha'\in\mathcal{V}}\frac{\sum_\alpha p_\rho(\alpha)q_{g\cdot\alpha,I}(\alpha',r)}{\sum_\alpha p_\rho(\alpha)Q_I(r\;|\;g\cdot\alpha)}A_{\alpha'}.
	\end{align}
	This agrees with the postmeasurement state predicted by quantum mechanics.  Therefore, the classical simulation algorithm correctly reproduces the outcome probabilities and the postmeasurement state predicted by quantum mechanics for a single layer of a QCM circuit.
	
	Now let $\rho(t)$ denote the state after $t$ layers of the circuit.  Then the argument above shows that the classical simulation algorithm correctly reproduces the Born rule probabilities $P_{\rho_0,g_{t+1},I_{t+1}}(r_{t+1}\;|\;r_1,r_2,\dots,r_t)$ as well as the postmeasurement state $\rho(t+1)$.  Therefore, by induction, the simulation algorithm correctly reproduces the outcome probabilities predicted by the Born rule for any QCM circuit. $\Box$

	\section{Partial characterization of vertices of \texorpdfstring{$\Lambda$}{Lambda}}\label{Section:CNCVertices}
	
	In Ref.~\cite{RaussendorfZurel2020} a classical simulation algorithm for quantum computation with magic states is introduced based on a quasiprobability representation.  Points in the generalized phase space over which the quasiprobability function is defined are associated with pairs $(\Omega,\gamma)$ where $\Omega\subset E$ is a cnc set and $\gamma:\Omega\rightarrow\mathbb{Z}_d$ is a noncontextual value assignment for $\Omega$.  For each point in phase space there is a corresponding phase space point operator defined as
	\begin{equation}\label{eq:CncPhasePointOperator}
		A_\Omega^\gamma=\frac{1}{d^n}\sum\limits_{b\in\Omega}\omega^{-\gamma(b)}T_b.
	\end{equation}
	For qubits, if $\Omega$ is a maximal cnc set then phase point operators $A_\Omega^\gamma$ of the form eq.~\eqref{eq:CncPhasePointOperator} are vertices of $\Lambda$~\cite{Heimendahl2019,ZurelRaussendorf2020}.  Vertices of the type eq.~\eqref{eq:CncPhasePointOperator} we call cnc vertices.
	
	A similar result holds for odd-prime-dimensional qudits.  Namely, phase point operators of the form $A_E^\gamma$ define facets of the stabilizer polytope~\cite{CormickGalvaoGottesman2006,VeitchEmerson2012}, or equivalently, by polar duality, $A_E^\gamma$ are vertices of $\Lambda$.  Here we show that this holds for qudits with any odd Hilbert space dimension.  This is the result of the following theorem.
	\begin{Theorem}\label{Theorem:CNCVertices}
		For any number $n$ of qudits with any odd Hilbert space dimension $d$, phase space point operators of the form
		\begin{equation}\label{eq:oddDPhasePointOperators}
			A_E^\gamma=\frac{1}{d^n}\sum\limits_{b\in E}\omega^{-\gamma(b)}T_b
		\end{equation}
		where $\gamma:E\longrightarrow\mathbb{Z}_d$ is a noncontextual value assignment for $E$ are vertices of $\Lambda$.
	\end{Theorem}
	This theorem provides a partial characterization of the vertices of $\Lambda$ which define the hidden variable model of Theorem~\ref{Theorem:HVMDef}. When $n\ge2$, the operators $A_E^\gamma$ are exactly the phase space point operators of Gross' Wigner function for odd-dimensional qudits~\cite{Gross2006,GrossPhD2008,DelfosseRaussendorf2017}, therefore, the standard multi-qudit phase point operators are a subset of the hidden variables of our model (when $n=1$, Gross' phase space point operators still have this form, but there are also other operators with this form~\cite{DelfosseRaussendorf2017}).  Note that the phase point operators of the form eq.~\eqref{eq:oddDPhasePointOperators} only exist when the qudit Hilbert space dimension $d$ is odd since noncontextual value assignments on $E$ exist only when $d$ is odd~\cite{OkayRaussendorf2017}.  In the case of qubits, some additional classes of vertices have been characterized~\cite{OkayRaussendorf2017}, but the vertices with the simplest description are the cnc type vertices of eq.~\eqref{eq:CncPhasePointOperator}.
	
	The proof of Theorem~\ref{Theorem:CNCVertices} requires the following lemma.
	
	\begin{Lemma}\label{Lemma:PauliExpMagnitude}
		For any vertex $A_\alpha$ of $\Lambda$ and any Pauli operator $T_a$, $|\Tr(T_aA_\alpha)|\le1$.
	\end{Lemma}
	
	\emph{Proof of Lemma~\ref{Lemma:PauliExpMagnitude}.} For any $\alpha\in\mathcal{V}$ and any $a\in E$,
	\begin{align}
		\Tr(T_aA_\alpha)=\sum\limits_{s\in\Gamma^{\langle a\rangle}}\Tr(\Pi_{\langle a\rangle}^sA_\alpha)\omega^{s(a)}
	\end{align}
	where $\Gamma^{\langle a\rangle}$ is the set of noncontextual value assignments on $\langle a\rangle$ and $\Pi_{\langle a\rangle}^s$ is the projector onto the eigenspace of the Pauli observable $T_a$ with eigenvalue $\omega^{s(a)}$.  $\{\Pi_{\langle a\rangle}^s|s\in\Gamma^{\langle a\rangle}\}$ is a projection-valued measure, i.e. $\sum_{s}\Pi_{\langle a\rangle}^s=\one$.  Therefore,
	\begin{align}
		\sum\limits_{s\in\Gamma^{\langle a\rangle}}\Tr(\Pi_{\langle a\rangle}^sA_\alpha)=&\Tr[\left(\sum\limits_{s\in\Gamma^{\langle a\rangle}}\Pi_{\langle a\rangle}^s\right)A_\alpha]=1
	\end{align}
	and so
	\begin{align}
		|\Tr(T_aA_\alpha)|\le&\sum\limits_{s\in\Gamma^{\langle a\rangle}}\left|\Tr(\Pi_{\langle a\rangle}^sA_\alpha)\omega^{s(a)}\right|\\
		\le&\sum\limits_{s\in\Gamma^{\langle a\rangle}}\left|\Tr(\Pi_{\langle a\rangle}^sA_\alpha)\right|\\
		=&\sum\limits_{s\in\Gamma^{\langle a\rangle}}\Tr(\Pi_{\langle a\rangle}^sA_\alpha)=1
	\end{align}
	which proves the claimed bound. $\Box$
	
	\bigskip
	
	We can now prove Theorem~\ref{Theorem:CNCVertices}.
	
	\emph{Proof of Theorem~\ref{Theorem:CNCVertices}.} First we need to show that the phase point operators of eq.~\eqref{eq:oddDPhasePointOperators} are in $\Lambda$.  This requires checking that the Hilbert-Schmidt inner product of $A_E^\gamma$ with the projector onto any stabilizer state is nonnegative.  For any maximal isotropic subgroup $I$ with any noncontextual value assignment $r:I\rightarrow\mathbb{Z}_d$,
	\begin{align}
		\Tr(\Pi_I^rA_E^\gamma)&=\frac{1}{d^{2n}}\sum\limits_{a\in I}\sum\limits_{b\in E}\omega^{-r(a)-\gamma(b)}\Tr(T_aT_b)\\
		&=\frac{1}{d^n}\sum\limits_{a\in I}\omega^{-r(a)-\gamma(-a)}\\
		&=\frac{1}{d^n}\sum\limits_{a\in I}\omega^{-r(a)+\gamma(a)}.
	\end{align}
	To obtain the last equality, note that with the phase convention chosen in eq.~\eqref{eq:PhaseConvention} we have $T_{-a}=T_a^\dagger=T_a^{-1}$ and $T_{0}=\one$. Therefore, $T_aT_{-a}=T_aT_a^{-1}=\one=T_{a+(-a)}$ and so by definition $\beta(a,-a)=0$. Further, eq.~\eqref{eq:noncontextuality condition} applied to the case $a=b=0$ implies $\gamma(0)=-\beta(0,0)=0$. Thus, with eq.~\eqref{eq:noncontextuality condition}, we have $\gamma(a)+\gamma(-a)-\gamma(0)=-\beta(a,-a)$, and so $\gamma(-a)=-\gamma(a)$.
	
	Here we have two cases.
	
	(I)~If $r|_I=\gamma|_I$ then we have
	\begin{equation}
		\Tr(\Pi_I^rA_\Omega^\gamma)=\frac{1}{d^n}\sum\limits_{a\in I}\omega^{-r(a)+\gamma(a)}=\frac{|I|}{d^n}=1.
	\end{equation}
	since $|I|=d^n$ by \cite[Theorem~1]{gheorghiu2014standard}.
	
	(II)~If $r|_I\ne\gamma|_I$ then by orthogonality of twisted characters~\cite{Cheng2015}, we have
	\begin{equation}
		\Tr(\Pi_I^rA_E^\gamma)=\frac{1}{d^n}\sum\limits_{a\in I}\omega^{-r(a)+\gamma(a)}=0.
	\end{equation}
	
	Therefore, in both cases $\Tr(\Pi_I^rA_E^\gamma)\ge0$.  This proves that the phase point operators of eq.~\eqref{eq:oddDPhasePointOperators} are in $\Lambda$.
	
	Since $A_E^\gamma\in\Lambda$, there exists an expansion of $A_E^\gamma$ as a convex combination of the vertices of $\Lambda$:
	\begin{equation}\label{eq:PPExpansion}
		A_E^\gamma=\sum\limits_{\alpha\in\mathcal{V}}p(\alpha)A_\alpha.
	\end{equation}
	Consider a Pauli operator $T_a,\; a\in E$.  We have $\Tr(T_aA_E^\gamma)=\omega^{\gamma(a)}$ so multiplying eq.~\eqref{eq:PPExpansion} by $T_a$ and taking a trace we get
	\begin{align}
		\omega^{\gamma(a)}=\sum\limits_{\alpha\in\mathcal{V}}p(\alpha)\Tr(T_aA_\alpha)
	\end{align}
	Taking the absolute value of this equation we get on the left hand side $\left|\omega^{\gamma(a)}\right|=1$ and on the right hand side
	\begin{align}
		\left|\sum\limits_{\alpha\in\mathcal{V}}p(\alpha)\Tr(T_aA_\alpha)\right|\le&\sum\limits_{\alpha\in\mathcal{V}}p(\alpha)|\Tr(T_aA_\alpha)|\\
		\le&\sum\limits_{\alpha\in\mathcal{V}}p(\alpha)=1.
	\end{align}
	Here the first inequality is the triangle inequality and the second inequality follows from Lemma~\ref{Lemma:PauliExpMagnitude}.  The second inequality is strict if $|\Tr(T_aA_\alpha)|<1$ for any $A_\alpha$ with $p(\alpha)>0$.  If this were the case then this would lead to a contradiction: $1<1$.  Therefore, $|\Tr(T_aA_\alpha)|=1$ for all $A_\alpha$ with $p(\alpha)>0$.
	
	Now consider the real part of the equation above:
	\begin{equation}
		\text{Re}\left[\omega^{\gamma(a)}\right]=\sum\limits_{\alpha\in\mathcal{V}}p(\alpha)\text{Re}\left[\Tr(T_aA_\alpha)\right].
	\end{equation}
	Since the coefficients are nonnegative and sum to one, this implies that $\text{Re}\left[\Tr(T_aA_\alpha)\right]=\text{Re}\left[\omega^{\gamma(a)}\right]$ for every $\alpha$ with $p(\alpha)>0$.  The same argument holds for the imaginary part.  Thus, $\Tr(T_aA_\alpha)=\omega^{\gamma(a)}$ for all $\alpha$ with $p(\alpha)>0$.  I.e. each $A_\alpha$ that appears with nonzero weight in the expansion of $A_E^\gamma$ in eq.~\eqref{eq:PPExpansion} must agree with $A_E^\gamma$ on the expectation of $T_a$ for all $a\in E$.  There is exactly one such operator in $\Herm_1(\mathcal{H})$, namely $A_E^\gamma$.  Therefore, $A_E^\gamma$ is a vertex of $\Lambda$. $\Box$

	\section{Mapping vertices of \texorpdfstring{$\Lambda_{m}$}{Lambda\_m} to \texorpdfstring{$\Lambda_{n}$}{Lambda\_n}}\label{Section:PhiMap}
	
	In this section we introduce a version of the $\Phi$-map \cite{OkayRaussendorf2021}, that embeds the $m$-qudit polytope $\Lambda_m$ as a subpolytope of the $n$-qudit polytope $\Lambda_n$ where $n\geq m$, that works over qudits for an arbitrary $d\geq 2$. However, for $d\neq 2$ this map fails to map vertices of $\Lambda_m$ to vertices of $\Lambda_n$.
	
	We will regard $E_m$ as a subgroup of $E_n$ by identifying it with $\Span{x_1,\cdots,x_m,z_1,\cdots,z_m}$. We will write $E_{n-m}^{(n)}$ for the subgroup $\Span{x_{m+1},\cdots,x_{n},z_{m+1},\cdots,z_{n}}$. These two subgroups intersect at the zero element and they generate the whole group. In this decomposition $E_{n-m}$ is identified with $E_{n-m}^{(n)}$ via the map $x_i\mapsto x_{m+i}$ and $z_i\mapsto z_{m+i}$. Any isotropic subgroup $J\subset E_{n-m}$ will be identified with its image in  $E_{n-m}^{(n)}$.
	
	\begin{Theorem}\label{Theorem:Phi-map}
		Let $\Pi_J^r$ denote an $(n-m)$-qudit stabilizer projector and $g$ denote an $n$-qudit Clifford unitary. The linear map
		\begin{equation}
			\Phi_{g,J}^r: \Herm\left(\mathbb{C}^{d^m}\right) \to \Herm\left(\mathbb{C}^{d^n}\right),\;\;\; X\mapsto  g( X\otimes \Pi_{J}^r )g^\dagger 
		\end{equation}
		is injective and sends $\Lambda_m$ to the subpolytope given by
		\begin{equation}
			\set{(\Pi_{J'}^{r'}Y\Pi_{J'}^{r'})/Tr(Y\Pi_{J'}^{r'})|\;Y\in \Lambda_n \text{ and }\Tr(Y\Pi_{J'}^{r'})\neq 0 }
		\end{equation}
		where $\Pi_{J'}^{r'}=g(\one \otimes \Pi_{J}^r) g^\dagger$.
	\end{Theorem}
	
	This partially generalizes to qudits of arbitrary dimension a stronger result which applies only to the case $d=2$~\cite{OkayRaussendorf2017}.
	
	\emph{Proof of Theorem~\ref{Theorem:Phi-map}.} To be able to distinguish Pauli operators we will write $T_a^{(n)}$ to indicate an $n$-qudit Pauli operator. With the choice of phase function $\phi$ in eq.~\eqref{eq:PauliOperators} we have (1) $T_0=\one$, (2) $\beta(a,ka)=0$ for all $a\in E_n$ and $k\in \Z_d$ and (3) $\beta(a,b)=0$ for all $a\in E_{m}$ and $b\in E_{n-m}^{(n)}$. It suffices to prove the theorem for the case where $(g,J,r)$ is given by $(\one,J_0,r_0)$. Here $\one$ is the identity operator, $J_0=\Span{x_{m+1},x_{m+2},\cdots,x_n}$, and $r_0$ is the value assignment defined by $r_0(x_i)=0$ for all $1\leq i\leq n-m$. This is because we can write
	\begin{align}
		\Phi_{g,J}^r(X) & = g \Phi_{\one,J}^{r}(X)  g^\dagger \\
		&= g (\one \otimes W) \Phi_{\one,J_0}^{r_0}(X) (\one \otimes W^\dagger)  g^\dagger \\
		& = V \Phi_{\one,J_0}^{r_0}(X) V^\dagger
	\end{align}
	where $W$ is the Clifford unitary such that $W \Pi_{J_0}^{r_0} W^\dagger =\Pi_{J}^{r} $ and $V = g (\one \otimes W)$. For the rest we will take $(g,J,r)=(\one,J_0,r_0)$. Let us write $\Phi=\Phi_{\one,J_0}^{r_0}$ for simplicity of notation.

	The map $\Phi$ sends $X$ to the tensor product $X\otimes \Pi_{J_0}^{r_0}$, and therefore it is linear because of the properties of the tensor product operation. For $X\in \Herm(\mathbb{C}^{d^m})$ we can write
	\begin{equation}\label{eq:X}
		X = \frac{1}{d^{m}} \sum_{a\in E_{m}} \alpha_a  T_a^{(m)}
	\end{equation} 
	which gives
	\begin{equation}\label{eq:Phi-U=1}
		\Phi(X) = X\otimes \Pi_{J}^r = \frac{1}{d^n} \sum_{a+b\in E_{m}+ J} \alpha_a   T_{a+b}^{(n)}.
	\end{equation}  
	In particular, 	$\Phi(X)=0$ implies that $X=0$, hence  $\Phi$ is injective. Next we show that $\Phi$ maps $\Lambda_m$ into $\Lambda_n$. First note that  $\Tr(\Phi(X)) = \alpha_0=1$ since $\Tr(X)=1$. For a maximal isotropic subgroup $I\subset E_n$ we compute 
	\begin{align}
		\Tr(\Pi_I^s\Phi(X))&=\frac{1}{d^{2n}} \sum_{c\in I}\sum_{a+b\in E_{m}+ J}
		\alpha_a\omega^{s(c)}\underbrace{\Tr(T_c^\dagger T_{a+b} )}_{d^n\delta_{c,a+b}}  \\
		&=\frac{1}{d^{n}}\sum_{a+b\in(E_{m}+J)\cap I}\alpha_a\omega^{s(a+b)}\\  
		&=\frac{|{K}\cap J|}{d^n} \sum_{a\in {K}\cap E_{m}} \alpha_a \omega^{s(a)} \left( \frac{1}{|{K}\cap J|} \sum_{b\in {K}\cap J} \omega^{s(b)} \right)\\
		&=\delta_{r|_{{K}\cap J},s|_{{K}\cap J}}\frac{|{K}\cap J|}{d^n}\sum_{a\in{K}\cap E_{m}}\alpha_a\omega^{s(a)}\\ 
		&=\delta_{r|_{{K}\cap J},s|_{{K}\cap J}}\frac{|{K}|}{d^n}\Tr(\Pi_{{K}\cap E_{m},s|_{{K}\cap E_{m}}}X)\geq0
	\end{align}
	where $K=(E_m+J)\cap I$. Therefore  $\Phi(X)\in \Lambda_n$.  This image is a convex polytope since the image of a convex polytope under a linear map is also a convex polytope. We want to show that the image of $\Phi$ is given by 
	\begin{equation}\label{eq:set-proj-conj}
		\set{\,\Pi_J^r Y \Pi_J^r/\Tr(Y \Pi_J^r)\,|\, Y\in \Lambda_n\;\text{ and }\;\Tr(Y \Pi_J^r)\neq 0 }.
	\end{equation}
	First observe that this set is contained in $\Lambda_n$ by Theorem~\ref{Theorem:HVMDef}. Writing $Y$  in the Pauli basis  $Y=(\sum_{c\in E_n} \beta_c T_c^{(n)})/d^n$ we obtain
	\begin{align}
		\Pi_{J}^{r} Y \Pi_{J}^{r} & = \frac{1}{d^n} \sum_{c\in E_n} \beta_c  \Pi_{J}^{r} T_c^{(n)}  \Pi_{J}^{r}\\
		& = \frac{1}{d^n} \sum_{a+b\in E_m+J} \beta_{a+b}  \Pi_{J}^{r} T_{a+b}^{(n)}  \\
		& = \frac{1}{d^n} \sum_{a+b\in E_m+J} \beta_{a+b}  \Pi_{J}^{r} (T_{a}^{(m)}\otimes  T_{b}^{(n-m)})  \\
		& = \frac{1}{d^n} \sum_{a+b\in E_m+J} \beta_{a+b}    T_{a}^{(m)}\otimes \Pi_{J}^{r}  \\
		& = \frac{1}{d^m} \sum_{a\in E_m} \left(\frac{1}{d^{n-m}}\sum_{b\in J} \beta_{a+b} \right)   T_{a}^{(m)}\otimes \Pi_{J}^{r}.
	\end{align}
	Defining $X$ by setting $\alpha_a=\Tr(Y\Pi_J^r)(\sum_{b\in J} \beta_{a+b})/d^{n-m}$ in eq.~\eqref{eq:X} this computation shows that $\Phi(X)=(\Pi_J^rY\Pi_J^r)/\Tr(Y\Pi_J^r)$. Moreover, $X$ belongs to $\Lambda_m$ since for any isotropic subgroup $I'\subset E_m$ and a value assignment $s'$ defined on it we have that $\Tr(X\Pi_{I'}^{s'})=\Tr((X\Pi_{I'}^{s'})\otimes  \Pi_J^r) = \Tr(\Phi(X) (\Pi_{I'}^{s'} \otimes \Pi_J^r))\geq 0 $ since $\Pi_{I'}^{s'} \otimes \Pi_J^r$ is a stabilizer projector and $\Phi(X)\in \Lambda_n$. This shows that the set in eq.~\eqref{eq:set-proj-conj} is contained in the image of $\Phi$. For the converse, we define $Y$ by setting $\beta_{a+b}=\alpha_a$ for all $b\in J$. Then by the computation above we find that $\Pi_J^r Y \Pi_J^r = \Phi(X)$ with $\Tr(Y\Pi_J^r)=\Tr(\Phi(X))=1$.
	$\Box$

	In contrast to the qubit case \cite{OkayRaussendorf2021}, the map $ \Phi $ does not necessarily map a vertex of $ \Lambda_m $ to a vertex of $ \Lambda_n $. For example, this is the case for odd qudit dimension $ d $ vertices of the form $ A_{E_m}^\gamma $, as in \eqref{eq:oddDPhasePointOperators}. 

	To see this, consider $ \Phi_{g,J}^r $ with $ g = \one $. Analogously to Lemma~\ref{lemma:SumsOfStabilizer CodeProjectors} in Appendix~\ref{Appendix:Stabilizer}, the operator 
	\begin{equation}
		\Phi_{\one, J}^{r}(A_{E_m}^\gamma) = A_{E_m}^\gamma \otimes \Pi_{J}^{r}
	\end{equation}
	can be written as a proper convex combination of vertices of $ \Lambda_n $:
	\begin{align}
		A_{E_m}^\gamma \otimes \Pi_{J}^{r} =& A_{E_m}^\gamma \otimes \left(\frac{1}{\left|\Gamma_{J,r}^{E_{n-m}}\right|} \sum_{\gamma' \in \Gamma_{J,r}^{E_{n-m}}} A_{E_{n-m}}^{\gamma'} \right)\\
		=&\frac{1}{\left|\Gamma_{J,r}^{E_{n-m}}\right|} \sum_{\gamma' \in \Gamma_{J,r}^{E_{n-m}}} A_{E_m}^\gamma \otimes A_{E_{n-m}}^{\gamma'}\\
		=&\frac{1}{\left|\Gamma_{J,r}^{E_{n-m}}\right|} \sum_{\gamma' \in \Gamma_{J,r}^{E_{n-m}}} A_{E_n}^{\gamma \star \gamma'},
	\end{align}
	where $ \Gamma_{J,r}^{E_{n-m}}  $ is the set of noncontextual value assignments $ r $ on $ E_{n-m} $ such that $ r_{|J} = r $ and $ \gamma \star \gamma_0 $ is the unique value assignment on $ E_n $ satisfying $ (\gamma \star \gamma_0)_{|E_m} = \gamma $ and $ (\gamma \star \gamma_0)_{|E_{n-m}} = \gamma_0 $.

	\section{Discussion}\label{Section:Discussion}
	
	In this paper, we have presented a hidden variable model which represents all components of quantum computation with magic states using only classical (nonnegative) probabilities.  In this model, magic states are represented by a probability distribution over a finite set according to eq.~\eqref{eq:HVMStateExpansion}.  Clifford gates are represented by a deterministic update rule---a map from the set of hidden variables to itself.  Pauli measurements are represented by a probabilistic update rule---a map from hidden variables to probability distributions over the set of hidden variables according to eq.~\eqref{eq:HVMPauliUpdate}.  This model is similar in form to many previously defined quasiprobability representations of quantum computation with magic states~\cite{Gross2006,Gross20062,GrossPhD2008,VeitchEmerson2012,DelfosseRaussendorf2015,HowardCampbell2017,RaussendorfZurel2020}, but with the distinguishing feature that in our model all states can be represented by a probability distribution.  No ``negative probabilities'' are required.  This is the generalization of the hidden variable model of Ref.~\cite{ZurelRaussendorf2020} to qudits of arbitrary local Hilbert space dimension.
	
	Since everything is represented probabilistically, the model leads to a classical simulation algorithm for quantum computation with magic states based on sampling from the defining probability distributions.  This is Algorithm~\ref{Algorithm:simAlg} presented in Section~\ref{Section:SimAlg}.  This algorithm is similar in structure to simulation algorithms based on sampling from quasiprobability distributions like the discrete Wigner function~\cite{VeitchEmerson2012,DelfosseRaussendorf2015,HowardCampbell2017,RaussendorfZurel2020,Zurel2020}, except that those algorithms are limited in their scope.  They can only simulate quantum circuits for which the input state of the circuit is represented by a probability distribution.  Since in our model any state can be represented by a probability distribution, we have no such limitation.  Note that although Algorithm~\ref{Algorithm:simAlg} can simulate any quantum computation, we make no claims that the simulation is efficient in general.  In fact, if a quantum computational speedup over classical computation is possible at all, as many believe, then efficient classical simulation must fail.
	
	There are, however, some important cases where the simulation is efficient.  For example, in the qubit case, it is known that the (efficient) classical simulation algorithm of Ref.~\cite{RaussendorfZurel2020} is a special case of this more general model/simulation algorithm.  This is a result of the fact that the phase point operators of eq.~\eqref{eq:CncPhasePointOperator} are vertices of $\Lambda$~\cite{Heimendahl2019,ZurelRaussendorf2020}, these are the cnc-type vertices.  In Section~\ref{Section:CNCVertices} we show that for higher-dimensional qudits as well there are vertices of $\Lambda$ of cnc-type.  We conjecture that the update of these vertices under Clifford gates and Pauli measurements will be efficiently computable classically. This would result in the simulation algorithm being efficient whenever the following two conditions are met: (i)~the support of the probability distribution representing the input state of the circuit is restricted to vertices of this type and (ii)~samples from the distribution representing the input state can be obtained efficiently.
	
	Of course not all vertices of $\Lambda$ are cnc-type.  For example, Ref.~\cite{Reichardt2009} characterizes all vertices of the two-qubit $\Lambda$ polytope.  Under the action of the Clifford group there are eight orbits of vertices, only two of which are cnc-type vertices.  Characterizing the remaining non-cnc-type vertices of $\Lambda$ for arbitrary $n$ and $d$ is an open problem which could expand the scope of efficiency of Algorithm~\ref{Algorithm:simAlg}.  This has already been partially achieved in the qubits case: Ref.~\cite{OkayRaussendorf2021} provides an efficient description, along with update rules under Pauli measurements, for one of the non-cnc-type orbits of the two qubit $\Lambda$ polytope.  As a result of the $\Phi$-map, which maps vertices of the $m$-qubit polytope to vertices of the $n$-qubit polytope with $m<n$, this also provides a description of a non-cnc-type orbit of vertices for the $\Lambda$ polytope on any number of qubits.  Ref.~\cite{OkayRaussendorf2021} also provides a reduction of the simulation algorithm that shows that the $\Phi$-map does not significantly increase the cost of classical simulation.  Therefore, the scope of efficiency of Algorithm~\ref{Algorithm:simAlg} in the qubit case goes beyond states supported on cnc-type vertices.  In section~\ref{Section:PhiMap} we show that a version of the $\Phi$-map also holds for qudits of arbitrary dimension.  We conjecture that a version of the corresponding reduction of the simulation algorithm also holds in this more general setting.  This would align the model and simulation algorithm with the resource theory perspective in which preparation of stabilizer states is considered a ``free'' operation, along with Clifford gates and Pauli measurements~\cite{VeitchEmerson2014}.
	
	We have seen evidence that the $\Lambda$ polytopes could prove to be of independent interest.  A subset of vertices, namely the cnc-type vertices, have already found a handful of applications~\cite{Heimendahl2019,KirbyLove2020,RaussendorfZurel2020}.  The remaining vertices are less well studied but have proven useful in certain areas as well~\cite{Reichardt2009,OkayRaussendorf2021}.  Therefore, we conclude by proposing the family of $\Lambda$ polytopes for arbitrary $n$ and $d$ as a subject of further study.
	
	The main open questions regarding the $\Lambda$ polytopes are ``where does the efficiency of classically simulating quantum computation end?'' (assuming it does), and now that we have found it's not always negativity in a quasiprobability representation, which physical property is responsible for the breakdown of efficient classical simulation?''.  In this paper we have shown that those questions do not only arise in the multi-qubit case, but rather in quantum computation on qudits of any dimension.

	\section*{Acknowledgements}
	
	This research was undertaken thanks, in part, to funding from the Max Planck-UBC-UTokyo Center for Quantum Materials and the Canada First Research Excellence Fund, Quantum Materials and Future Technologies Program.  M.Z. and R.R. are funded by the National Science and Engineering Research Council of Canada.  C.O. is supported by the Air Force Office of Scientific Research under award number FA9550-21-1-0002.  A.H. is partially supported by the Deutsche Forschungsgemeinschaft (DFG, German Research Foundation) under the Priority Program Compressed Sensing in Information Processing (CoSIP, project number SPP1798) and under Germany's Excellence Strategy -- Cluster of Excellence Matter and Light for Quantum Computing (ML4Q) EXC 2004/1 -- 390534769. This work was also supported by the Horizon Europe project FoQaCiA, GA no. 101070558.

	\printbibliography

	\appendix
	
	\section{The qudit stabilizer formalism}\label{Appendix:Stabilizer}
	
	The stabilizer formalism~\cite{Gottesman1997} describes a large family of quantum error correcting codes, as well as a broader framework for describing quantum error correction and fault-tolerant quantum computation.  In this appendix we review some features of the stabilizer formalism for systems of qudits~\cite{Gottesman1999oddDClifford} which are useful for some of the proofs in the main text.
	
	A stabilizer code is specified by a pair $(I,r)$ where $I$ is an isotropic subgroup of $E$, and $r:I\rightarrow\mathbb{Z}_d$ is a noncontextual value assignment for $I$.  The codespace of the code is the simultaneous $+1$-eigenspace of the Pauli operators $S_I^r=\{\omega^{-r(a)}T_a\;|\;a\in I\}$.  $S_I^r$ is called the stabilizer group of the code.  The projector onto this eigenspace, or equivalently, the projector onto the eigenspace of the Pauli observables labeled by $I$ corresponding to eigenvalues $\{\omega^{r(a)}\;|\;a\in I\}$ is
	\begin{equation}
		\Pi_I^r=\frac{1}{|I|}\sum\limits_{b\in I}\omega^{-r(b)}T_b.
	\end{equation}
	The dimension of this eigenspace is $d^n/|I|$~\cite{gheorghiu2014standard}.  In particular, we have the following lemma.
	\begin{Lemma}\label{Lemma:OrderMaxIso}
		The order of a maximal isotropic subgroup of $E$ is $d^n$.
	\end{Lemma}
	For the proof of Lemma~\ref{Lemma:OrderMaxIso} see Theorem~1 in Ref.~\cite{gheorghiu2014standard}.  If $I$ is a maximal isotropic subgroup of $E$, $|I|=d^n$, and there is a unique quantum state fixed by a stabilizer group $S_I^r$.  Such states are called stabilizer states.
	
	Stabilizer code projectors can be constructed from products of stabilizer code projectors of higher rank.  If $\{I_k\;|\; I_k\subset I\}$ are such that $I=\bigcup_{k}I_k$, and the value assignments $r_k:I_k\rightarrow\mathbb{Z}_d$ satisfy $r_k= r|_{I_k}$, then
	\begin{equation}
		\Pi_I^r=\prod_{k}\Pi_{I_k}^{r_k}.
	\end{equation}
	Stabilizer measurements can also be coarse-grained to give stabilizer projectors of higher rank.  This is the result of the following lemma.
	\begin{Lemma}\label{lemma:SumsOfStabilizer CodeProjectors}
		If $ I $ is a non-maximal isotropic subgroup of $E$ and $\Pi_I^r$ is a stabilizer code projector for some noncontextual value assignment $r: I \to \Z_d$, then for any isotropic subgroup $I'$ containing $I$
		\begin{equation}\label{eq:summing-ncva}
			\sum_{r' \in \Gamma_{I,r}^{I'}} \Pi_{I'}^{r'} = \Pi_I^r,
		\end{equation}
		where $\Gamma_{I,r}^{I'}$ is the set of all noncontextual value assignments on $I'$ such that $r'|_{I} = r$.
	\end{Lemma}
	
	\emph{Proof of Lemma~\ref{lemma:SumsOfStabilizer CodeProjectors}.} The proof is obtained by adapting the proof of Ref.~\cite[Lemma~1]{RaussendorfZurel2020}.  Let $I,I'$ be isotropic subgroups of $E$ such that $I\subsetneq I'$.  Let $r:I\rightarrow\mathbb{Z}_d$ be a noncontextual value assignment for $I$ and $\Gamma_{I,r}^{I'}$ be the set of all noncontextual value assignments on $I'$ satisfying
	\begin{equation}
		r'|_I=r\quad\forall r'\in\Gamma_{I,r}^{I'}.
	\end{equation}
	Then
	\begin{align}
		\sum\limits_{r'\in\Gamma_{I,r}^{I'}}\Pi_{I'}^{r'}=&\sum\limits_{r'\in\Gamma_{I,r}^{I'}}\frac{1}{|I'|}\sum\limits_{a\in I'}\omega^{-r'(a)}T_a\\
		=&\frac{1}{|I'|}\sum\limits_{a\in I'}\left[\sum\limits_{r'\in\Gamma_{I,r}^{I'}}\omega^{-r'(a)}\right]T_a.
	\end{align}
	We have two cases for the inner sum in the last expression.  If $a\in I$ then $r'(a)=r(a)$ for all $r'\in\Gamma_{I,r}^{I'}$.  Therefore,
	\begin{equation}
		\sum\limits_{r'\in\Gamma_{I,r}^{I'}}\omega^{-r'(a)}=\left|\Gamma_{I,r}^{I'}\right|\omega^{-r(a)}.
	\end{equation}
	In the second case, $a\notin I$.  $\Gamma_{I,r}^{I'}$ is the coset of a vector space, the proof of this is analogous to the proof of Ref.~\cite[Lemma~2]{RaussendorfZurel2020} which applies only to qubits.  Therefore, by character orthogonality,
	\begin{equation}
		\sum\limits_{r'\in\Gamma_{I,r}^{I'}}\omega^{-r'(a)}=0.
	\end{equation}
	Thus,
	\begin{align}
		\sum\limits_{r'\in\Gamma_{I,r}^{I'}}\Pi_{I'}^{r'}=&\frac{1}{|I'|}\sum\limits_{a\in I'}\left|\Gamma_{I,r}^{I'}\right|\delta_{a\in I}\omega^{-r(a)}T_a\\
		=&\frac{\left|\Gamma_{I,r}^{I'}\right|}{|I'|}\sum\limits_{a\in I}\omega^{-r(a)}T_a.
	\end{align}
	We have $|\Gamma_{I,r}^{I'}|=|I'|/|I|$.  Therefore,
	\begin{equation}
		\sum\limits_{r'\in\Gamma_{I,r}^{I'}}\Pi_{I'}^{r'}=\frac{1}{|I|}\sum\limits_{a\in I}\omega^{-r(a)}T_a=\Pi_I^r
	\end{equation}
	which proves the lemma. $\Box$
	
	\begin{Corollary}\label{cor:stab-polytope-as-convex-hull-of-proj}
		The stabilizer polytope (the convex hull of stabilizer states) is
		\begin{equation}\label{eq:alternative-description-SP}
			\SP = \textup{conv} \left  \{ \frac{|I|}{d^n}\Pi_I^r \; | \; \Pi_I^r \textup{ stabilizer code projector} \right \}.
		\end{equation}
	\end{Corollary}
	
	\medskip
	
	The above lemmas describe structural properties of stabilizer code projectors.  It will also be useful for us to describe how stabilizer projectors behave under the dynamical operations of quantum computation with magic states---Clifford gates and Pauli measurements.  First, we have the following result regarding Clifford gates.
	\begin{Lemma}\label{Lemma:StabilizerClifford}
		For any Clifford group element $g\in\Cl$, and any stabilizer state $\ket{\sigma}\in\mathcal{S}$,
		\begin{equation}
			g\ket{\sigma}\bra{\sigma}g^\dagger=\ket{\sigma'}\bra{\sigma'}
		\end{equation}
		where $\ket{\sigma'}\in\mathcal{S}$ is a stabilizer state.  I.e. Clifford group operations map stabilizer states to stabilizer states.
	\end{Lemma}
	
	\emph{Proof of Lemma~\ref{Lemma:StabilizerClifford}.} The action of the Clifford group on the Pauli operators is defined in eq.~\eqref{eq:CliffordAction}.  With this equation, for a projector onto a stabilizer state $\ket{\sigma}$ corresponding to maximal isotropic subgroup $I\subset E$ and noncontextual value assignment $r:I\rightarrow\mathbb{Z}_d$ we have
	\begin{align}
		g\ket{\sigma}\bra{\sigma}g^\dagger=&g\Pi_I^rg^\dagger=\frac{1}{|I|}\sum\limits_{a\in I}\omega^{-r(a)}gT_ag^\dagger\\
		=&\frac{1}{|I|}\sum\limits_{a\in I}\omega^{-r(a)+\Phi_g(a)}T_{S_g(a)}\\
		=&\frac{1}{|I|}\sum\limits_{a\in g\cdot I}\omega^{-g\cdot r(a)}T_a=\Pi_{g\cdot I}^{g\cdot r}
	\end{align}
	where $g\cdot I=\{S_g(a)\;|\;a\in I\}$ and $g\cdot r$ is defined by the relation
	\begin{equation}
		g\cdot r(S_g(a))=r(a)-\Phi_g(a)\quad\forall a\in I.
	\end{equation}
	In order to show that $g\ket{\sigma}\bra{\sigma}g^\dagger$ is a projector onto a stabilizer state it suffices to show that $g\cdot I$ is a maximal isotropic subgroup of $E$ and that $g\cdot r:g\cdot I\rightarrow\mathbb{Z}_d$ is a noncontextual value assignment for $g\cdot I$.
	
	$g\cdot I$ is isotropic since $I$ is isotropic and $S_g$ is a symplectic operation.  Also, $|g\cdot I|=|I|=d^n$ so by Lemma~\ref{Lemma:OrderMaxIso}, $g\cdot I$ is a maximal isotropic subgroup.
	
	Since $T_0\propto\one$
	\begin{equation}
		gT_0g^\dagger=T_0
	\end{equation}
	so $\Phi_g(0)=0$ for any $g\in\Cl$.  Therefore,
	\begin{equation}
		\omega^{-g\cdot r(0)}T_0=\omega^{-r(0)+\Phi_g(0)}T_0=\omega^{-r(0)}T_0=\one.
	\end{equation}
	For any $a,b\in g\cdot I$, there exist $c,d\in I$ such that $S_g(c)=a$ and $S_g(d)=b$.  Computing the product $gT_cT_dg^\dagger$ in two different ways we have
	\begin{align}
		gT_cT_dg^\dagger=&gT_cg^\dagger gT_dg^\dagger\\
		=&\omega^{\Phi_g(c)+\Phi_g(d)}T_{S_g(c)}T_{S_g(d)}\\
		=&\omega^{\Phi_g(c)+\Phi_g(d)-\beta(a,b)}T_{a+b}
	\end{align}
	and
	\begin{align}
		gT_cT_dg^\dagger=&\omega^{-\beta(c,d)}gT_{c+d}g^\dagger\\
		=&\omega^{-\beta(c,d)+\Phi_g(c+d)}T_{a+b}.
	\end{align}
	Therefore, $-\Phi_g(c)-\Phi_g(d)+\Phi_g(c+d)=\beta(c,d)-\beta(a,b)$ and so
	\begin{align}
		&g\cdot r(a)+g\cdot r(b)-g\cdot r(a+b)\\
		&=r(c)+r(d)-r(c+d)-\Phi_g(c)-\Phi_g(d)+\Phi_g(c+d)\\
		&=-\beta(a,b).
	\end{align}
	Thus, $g\cdot r$ satisfies eq.~\eqref{eq:noncontextuality condition} and $\Pi_{g\cdot I}^{g\cdot r}$ is a projector onto a stabilizer state. $\Box$
	
	\medskip
	
	The update of stabilizer states under Pauli measurements is probabilistic in general.  It is described in the following lemma.
	\begin{Lemma}\label{Lemma:StabilizerPauli}
		For any isotropic subgroups $I,J\subset E$ and any noncontextual value assignments $r:I\rightarrow\mathbb{Z}_d$ and $s:J\rightarrow\mathbb{Z}_d$,
		\begin{enumerate}
			\item if $r|_{I\cap J}=s|_{I\cap J}$ then
			\begin{equation}
				\Tr(\Pi_J^s\Pi_I^r)=\frac{|I\cap J|}{|I||J|}d^n>0
			\end{equation}
			and
			\begin{equation}
				\frac{\Pi_I^r\Pi_J^s\Pi_I^r}{\Tr(\Pi_J^s\Pi_I^r)}=\frac{|J\cap I^\perp|}{|J|}\Pi_{I+J\cap I^\perp}^{r\star s}
			\end{equation}
			where $r\star s$ is the unique noncontextual value assignment on the set $I+J\cap I^\perp$ such that $r\star s|_I=r$ and $r\star s|_{J\cap I^\perp}=s|_{J\cap I^\perp}$.
			\item If $r|_{I\cap J}\ne s|_{I\cap J}$ then
			\begin{equation}
				\Tr(\Pi_J^s\Pi_I^r)=0\quad\text{and}\quad\Pi_I^r\Pi_J^s\Pi_I^r=0.
			\end{equation}
		\end{enumerate}
	\end{Lemma}
	
	\emph{Proof of Lemma~\ref{Lemma:StabilizerPauli}.} Let $I,J\subset E$ be isotropic subgroups with noncontextual value assignments $r:I\rightarrow\mathbb{Z}_d$ and $s:J\rightarrow\mathbb{Z}_d$.
	
	\emph{Case 1: $r|_{I\cap J}=s|_{I\cap J}$.} Let $r\star s$ denote the unique noncontextual value assignment on the set $I+J\cap I^\perp$ such that $r\star s|_I=r$ and $r\star s|_{J\cap I^\perp}=s|_{J\cap I^\perp}$.  We calculate
	\begin{align}
		\Pi_I^r\Pi_J^s\Pi_I^r=&\frac{1}{|I|^2|J|}\sum\limits_{a,c\in I}\sum\limits_{b\in J}\omega^{-r(a)-r(c)-s(b)}T_aT_bT_c\\
		=&\frac{1}{|I|^2|J|}\sum\limits_{a,c\in I}\sum\limits_{b\in J}\omega^{-r(a)-r(c)-s(b)-\beta(a,c)+[b,c]}T_{a+c}T_b\\
		=&\frac{1}{|I|^2|J|}\sum\limits_{a,c\in I}\sum\limits_{b\in J}\omega^{-r(a+c)-s(b)+[b,c]}T_{a+c}T_b\\
		=&\frac{1}{|I||J|}\Pi_I^r\sum\limits_{b\in J}\omega^{-s(b)}\left[\sum\limits_{c\in I}\omega^{[b,c]}\right]T_b\\
		=&\frac{1}{|J|}\Pi_I^r\sum\limits_{b\in J\cap I^\perp}\omega^{-s(b)}T_b\\
		=&\frac{1}{|I||J|}\sum\limits_{a\in I}\sum\limits_{b\in J\cap I^\perp}\omega^{-r(a)-s(b)-\beta(a,b)}T_{a+b}\\
		=&\frac{1}{|I||J|}\sum\limits_{a\in I}\sum\limits_{b\in J\cap I^\perp}\omega^{-r\star s(a+b)}T_{a+b}.\label{eq:PauliCalculation}
	\end{align}
	Each Pauli operator in the above sum appears with the same multiplicty.  For an element $a\in I+J\cap I^\perp$, let $\mu(a)$ denote the number of distinct pairs $(b,c)\in I\times J\cap I^\perp$ such that $b+c=a$.  We have $\mu(a)=\mu(0)$ for any $a\in I+J\cap I^\perp$.  To see this, suppose $a\in I+J\cap I^\perp$ and let $(c_1,d_1),(c_2,d_2),\dots,(c_{\mu(a)},d_{\mu(a)})$ be the distinct pairs in $I\times J\cap I^\perp$ such that $c_j+d_j=a$.  Then the pairs $(c_j-c_1,d_j-d_1)\in I\times J\cap I^\perp$ for $j=2,\dots,\mu(a)$ together with the pair $(0,0)\in I\times J\cap I^\perp$ show that $\mu(0)\ge\mu(a)$.
	
	Now let $(c_1,d_1),(c_2,d_2),\dots,(c_{\mu(0)},d_{\mu(0)})$ denote the $\mu(0)$ distinct pairs in $I\times J\cap I^\perp$ such that $c_j+d_j=0$, and $(c,d)\in I\times J\cap I^\perp$ be such that $c+d=a$.  Then the pairs $(c_j+c,d_j+d),\;j=1,2,\dots,\mu(0)$ show that $\mu(a)\ge\mu(0)$.  Therefore, $\mu(a)=\mu(0)$ for any $a\in I+J\cap I^\perp$.  Here we can see $\mu(0)=|I\cap J|$
	
	Then we have
	\begin{align}
		\Pi_I^r\Pi_J^s\Pi_I^r=&\frac{1}{|I||J|}\sum\limits_{a\in I}\sum\limits_{b\in J\cap I^\perp}\omega^{-r\star s(a+b)}T_{a+b}\\
		=&\frac{\mu(0)}{|I||J|}\sum\limits_{a\in I+J\cap I^\perp}\omega^{-r\star s(a)}T_a\\
		=&\frac{|I\cap J||I+J\cap I^\perp|}{|I||J|}\Pi_{I+J\cap I^\perp}^{r\star s}.
	\end{align}
	Since $|I+J\cap I^\perp|=|I|\cdot|J\cap I^\perp|/|I\cap J|$,
	\begin{equation}
		\Pi_I^r\Pi_J^s\Pi_I^r=\frac{|J\cap I^\perp|}{|J|}\Pi_{I+J\cap I^\perp}^{r\star s}.
	\end{equation}
	Taking a trace of this equation we get
	\begin{align}
		\Tr(\Pi_I^r\Pi_J^s)=&\Tr(\Pi_I^r\Pi_J^s\Pi_I^r)\\
		=&\frac{|J\cap I^\perp|}{|J|}\Tr(\Pi_{I+J\cap I^\perp}^{r\star s})\\
		=&\frac{|J\cap I^\perp|}{|J|}\cdot\frac{d^n}{|I+J\cap I^\perp|}\\
		=&\frac{|I\cap J|}{|I||J|}d^n>0.
	\end{align}
	
	\emph{Case 2: $r|_{I\cap J}\ne s|_{I\cap J}$.}
	Denote by $S_J^s$ the stabilizer group $\{\omega^{-s(b)}T_b\;|\;b\in J\}$ and denote by $V_J^s$ the subspace of the Hilbert space $\mathbb{C}^{d^n}$ stabilized by $S_J^s$.  By assumption, there exists a $c\in I\cap J$ such that $r(c)\ne s(c)$.  Therefore, for any vector $\ket{\psi}\in V_J^s$,
	\begin{align}
		\Pi_I^r\ket{\psi}=&\Pi_I^r\omega^{-s(c)}T_c\ket{\psi}\\
		=&\frac{1}{d^n}\sum\limits_{a\in I}\omega^{-r(a)-s(c)}T_aT_c\ket{\psi}\\
		=&\frac{1}{d^n}\sum\limits_{a\in I}\omega^{-r(a)-s(c)-\beta(a,c)}T_{a+c}\ket{\psi}\\
		=&\frac{1}{d^n}\sum\limits_{a\in I}\omega^{-r(a+c)-r(c)-s(c)}T_{a+c}\ket{\psi}\\
		=&\frac{1}{d^n}\sum\limits_{a\in I}\omega^{-r(a)-r(c)-s(c)}T_a\ket{\psi}\\
		=&\omega^{-r(c)-s(c)}\Pi_I^s\ket{\psi}.
	\end{align}
	That is, $\Pi_I^s\ket{\psi}=\omega^{-r(c)-s(c)}\Pi_I^s\ket{\psi}$.  But since $r(c)\ne s(c)$, $\omega^{-r(c)-s(c)}\ne1$ so this relation can be true only if $\Pi_I^r\ket{\psi}=0$.
	
	We can write
	\begin{equation}
		\Pi_J^s=\sum_{i=1}^{\dim(V_J^s)}\ket{\psi_i}\bra{\psi_i}
	\end{equation}
	where $\{\ket{\psi_i}\;|\;1\le i\le\dim(V_J^s)\}$ is a basis for $V_J^s$.  Therefore,
	\begin{equation}
		\Pi_I^r\Pi_J^s\Pi_I^r=\sum\limits_{i=1}^{\dim(V_J^s)}\Pi_I^r\ket{\psi_i}\bra{\psi_i}\Pi_I^r=0.
	\end{equation}
	Taking a trace of this equation we get
	\begin{equation}
		\Tr(\Pi_I^r\Pi_J^s\Pi_I^r)=\Tr(\Pi_I^r\Pi_J^s)=0.
	\end{equation}
	This proves the lemma. $\Box$

	\section{Proof of Lemma~\ref{lemma:compactness-of-Lambda}}\label{Appendix:Lemma1}
	
	To prove the lemma, we will use the concept of polar duality for objects in the affine space $ \Herm_1(\mathcal{H}) $ (see Ref.~\cite{Ziegler1995} for a discussion of polar duality). For a Hilbert space $ \mathcal{H} $ with inner product $ \langle \cdot , \cdot \rangle $ define\footnote{The usual definition differs slightly by an irrelevant scaling factor.} the \emph{polar dual} of a set $ P \subset \mathcal{H}$ as 
	\begin{equation}
		P^* = \left\{  x \in \R^N \; \bigg| \; \langle x,y \rangle \ge -\frac{1}{d^n}  \textup{ for all } y \in P \right\}. 
	\end{equation}
	If $ P = \textup{conv} \{ v_1,\ldots, v_k \} $ is a polytope, then this simplifies to
	\begin{equation}
		P^*= \left\{  x \in \R^N \; \bigg| \; \langle x,v_i \rangle \ge -\frac{1}{d^n}, \; i, \ldots, k\right\}.
	\end{equation}
	If $ P \subset Q \subset \mathcal{H} $, then obviously $Q^* \subset P^*$.
	
	In our setting, we are interested in objects 
	living in the affine space of matrices of trace one $ \Herm_1(\mathcal{H}) $. We can project $ \Herm_1(\mathcal{H}) $ into the linear subspace of Hermitian matrices of trace zero $ \Herm_0(\mathcal{H}) $ via the transformation 
	\begin{equation}
		\pi: X\in\Herm_1(\mathcal{H}) \to \Herm_0(\mathcal{H}), \;  X \mapsto X - \frac{1}{d^n} \one. 
	\end{equation}
	Now, observe that for $ X,Y \in \Herm_1(\mathcal{H}) $ we have $\Tr(XY) \ge 0$ if and only if
	\begin{equation}
		\Tr(\pi(X)\pi(Y)) = \Tr(XY) - \Tr(X \frac{1}{d^n}\one) \ge -\frac{1}{d^n}.
	\end{equation} 
	Hence, by associating $ \Herm_0(\mathcal{H}) $ and $ \Herm_1(\mathcal{H}) $, we define for a set $ M \subset  \Herm_1(\mathcal{H}) $
	\begin{equation}\label{eq:Dual-in-affine-space}
		M^* := \{  X \in  \Herm_1(\mathcal{H})  \; | \;  \Tr(XY) \ge 0 \; \textup{ for all } Y \in M \}.
	\end{equation}
	If  $ P = \textup{conv} \{ X_1,\ldots, X_m \} \subset \Herm_1(\mathcal{H}) $ is a polytope, then define its \emph{polar dual} as 
	\begin{equation}
		P^* = \{ Y \in \Herm_1(\mathcal{H}) \, | \, \Tr(X_iY) \ge 0, \, i = 1,\ldots,m  \}.
	\end{equation}
	Thus, $ \Lambda= \SP^* $ for $ \SP $ being the \emph{stabilizer polytope}: $\SP := \textup{conv}\left\{ \ket{\sigma}\bra{\sigma} \; | \; \sigma \in \mathcal{S} \right\}$.
	
	To prove that the set $ \Lambda $ is bounded, it will suffice to show that $ \SP $ contains a set $ M $, whose dual $ M^* $ is bounded. Additionally, we will make us of the concept of \emph{dilation} \cite[Chap.~9] {SzarekAubrun2017}: for a set $ M \subset \Herm_1(\mathcal{H}) $ define its dilation centered at the maximally mixed state via
	\begin{equation}\label{eq:dilation-in-Herm-1}
		c\cdot M := \left   \{ \frac{1}{d^n} \one+ c \pi(X) \; | \; X \in M  \right \}
	\end{equation}
	where $\pi:\Herm_1(\mathcal{H})\rightarrow\Herm_0(\mathcal{H})$ is the projection that maps $X\in\Herm_1(\mathcal{H})$ to $X-\frac{1}{d^n}\one$.  The dilation has the following property: 
	\begin{Lemma}\label{lemma:dilation}
		The dilation of a set $ M \subset \Herm_1(\mathcal{H}) $ satisfies 
		\begin{align}
			(c \cdot M)^* = \frac{1}{c} \cdot M^*.
		\end{align}
	\end{Lemma}
	\emph{Proof of Lemma~\ref{lemma:dilation}.} Observe that every $ A,Y \in \Herm_1(\mathcal{H}) $ can be written as  
	\[Y=\frac{1}{d^n} \one+ c \pi(X), \quad A= \frac{1}{d^n} \one + \frac{1}{c} \pi(B) \] for suitable $ B,X \in \Herm_1(\mathcal{H}) $. 
	Then 
	\begin{align}
		&\Tr(AY)=\frac{1}{d^{2n}} \Tr(\one) + \Tr(\pi(B)\pi(X))\\
		&=\frac{1}{d^n} + \Tr(B(X-\frac{1}{d^n}\one))+\Tr(-\frac{1}{d^n}\one(X-\frac{1}{d^n}\one))\\
		&=\frac{1}{d^n} -\frac{1}{d^n}+ \Tr(BX) \\ & = \Tr(BX).
	\end{align}
	Hence, if $ Y \in c \cdot M$ with $ X \in M $, then $ A \in (c \cdot M)^* $ if and only if 
	\begin{align}
		0 \le \Tr(AY) = \Tr(BX) \quad \Longleftrightarrow \quad B \in M^*,
	\end{align} 
	which is equivalent to $ A \in \frac{1}{c} \cdot M^* $
	$ \Box $
	
	Having introduced all necessary concepts, we will proceed with the proof of Lemma~\ref{lemma:compactness-of-Lambda}.
	
	\emph{Proof of Lemma~\ref{lemma:compactness-of-Lambda}.}
	By definition, $ \Lambda $ is a polyhedron and therefore convex and closed.  To prove that $ \Lambda $ is bounded, the previous discussion implies that it suffices to find a set $ M \subset \SP $ such that $ M^* $ is bounded.
	The object we will choose here will be a dilation of the full-dimensional simplex 
	\begin{align}
		\vartriangle_{\Herm_1(\mathcal{H})}:=\textup{conv}\bigg\{& A_\gamma := \frac{1}{d^n}\sum_{u \in E} \omega^{\gamma(u)}T_u\;\bigg|\\
		&\gamma: E \to \Z_d, \;\gamma(u+v) = \gamma(u)+\gamma(v) \bigg\}.
	\end{align}
	The simplex $ \vartriangle_{\Herm_1(\mathcal{H})} $ is the Wigner simplex for $ d $ being an odd prime \cite{VeitchEmerson2012, VeitchEmerson2014}. For every $ d $, it is a full-dimensional polytope as the convex-hull of $ d^{2n} $ affinely independent vertices $ A_\gamma $ in the $ (d^{2n}-1) $-dimensional affine space $ \Herm_1(\mathcal{H}) $.  Due to $ \Tr(A_\gamma A_{\gamma'}) = \delta_{\gamma = \gamma'} $ for additive functions $ \gamma, \gamma': E \to \Z_d $, one can easily verify that the simplex $ \vartriangle_{\Herm_1(\mathcal{H})} $ has the following hyperplane description: 
	\begin{equation}
		\vartriangle_{\Herm_1(\mathcal{H})} = \{ X \in \Herm_1(\mathcal{H}) \; | \; \Tr(A_\gamma X ) \ge 0 \},
	\end{equation}
	which makes it a self-dual simplex, i.e.~$ \vartriangle_{\Herm_1(\mathcal{H})} = \vartriangle_{\Herm_1(\mathcal{H})}^* $. 
	
	Now, Lemma~\ref{lemma:dilation} implies that for $ c > 0 $ the simplex $ (c \cdot \vartriangle_{\Herm_1(\mathcal{H})})^* $ is bounded, since
	\begin{align}
		(c \cdot \vartriangle_{\Herm_1(\mathcal{H})})^* = \frac{1}{c} \vartriangle_{\Herm_1(\mathcal{H})}^* = \frac{1}{c} \vartriangle_{\Herm_1(\mathcal{H})}.
	\end{align}
	Hence, it suffices to show that $ c \cdot \vartriangle_{\Herm_1(\mathcal{H})} \subset \SP $ for some $ c > 0 $, because then
	\begin{align}
		\Lambda = \SP^* \subset (c \cdot \vartriangle_{\Herm_1(\mathcal{H})})^* = \frac{1}{c} \vartriangle_{\Herm_1(\mathcal{H})}
	\end{align}
	implies that $ \Lambda $ is bounded.  Therefore, we will show that dilations of the vertices of $ \vartriangle_{\Herm_1(\mathcal{H})} $ are contained in $ \SP $, i.e.there is $  c > 0$ such that 
	\begin{equation}
		\frac{1}{d^n} \one + c(A_\gamma -\frac{1}{d^n}\one) \in \SP
	\end{equation}
	for all additive maps $ \gamma $.
	
	To achieve this, we will write $ 1/d^n \one + c (A_\gamma - 1/d^n \one) $ as a convex combination of normalized projectors of the form $ \frac{|\langle a \rangle |}{d^n}\Pi_{\langle a\rangle}^r \in \SP$ for $ a \in E $ with noncontextual value assignments $ r: \langle a \rangle \to \Z_d $. Due to Corollary~\ref{cor:stab-polytope-as-convex-hull-of-proj} in Appendix~\ref{Appendix:Stabilizer}, the elements $ \frac{|\langle a \rangle |}{d^n}\Pi_{\langle a\rangle}^r $ are indeed contained in $ \SP $.
	A noncontextual value assignment on a line $ \langle a \rangle $ is always additive, that is 
	\begin{equation}\label{eq:additivity-of-ncva-on-groups-gen-by-one-element}
		r(ka) = kr(a), \quad\forall k \in \Z_d
	\end{equation}
	because with the phase convention chosen in eq.~\eqref{eq:PhaseConvention}, it can be checked directly that $T_aT_{ka}=T_{(k+1)a}$, and so by definition $ \beta(a,ka) = 0 $ for all $ a \in E $.
	Let $ C = \{a_1,\ldots, a_N\} \subset E $ be a set that ``covers'' $ E $, that is 
	\begin{equation}
		E = \cup_{a \in C} \langle a \rangle. 
	\end{equation}
	For each subset $ \mathcal{I} \subset [N] := \{1,\ldots,N\} $, there is an $ a_{\mathcal{I}} \in E $ such that 
	\begin{equation}
		\langle a_{\mathcal{I}}\rangle : = \bigcap_{k \in \mathcal{I}} \langle a_k \rangle.
	\end{equation}
	First, we will write $ A_\gamma $ as a linear combination of stabilizer code projectors $ \Pi_{\langle a\rangle }^r $.
	We claim that 
	\begin{align}\label{eq:Agamma-as-linear-combination}
		A_\gamma = \frac{1}{d^n}\sum_{\mathcal{I} \subset [N]} (-1)^{|\mathcal{I}|+1} \cdot   |\langle a_{\mathcal{I}} \rangle |\cdot  \Pi_{\langle a_{\mathcal{I}}\rangle }^{\gamma|_{\langle a_{\mathcal{I}} \rangle }}.
	\end{align}
	Since $ \gamma $ is additive, the restriction $ \gamma|_{\langle a_{\mathcal{I}} \rangle}: \langle a_{\mathcal{I}} \rangle \to \Z_d$ satisfies \eqref{eq:additivity-of-ncva-on-groups-gen-by-one-element} and defines a noncontextual value assignment on $ \langle a_{\mathcal{I}} \rangle $.
	We rewrite the right hand side of \eqref{eq:Agamma-as-linear-combination} in the following way: 
	\begin{align}
		&\frac{1}{d^n}\sum_{\mathcal{I}\subset [N]} (-1)^{|\mathcal{I}|+1} \cdot|\langle a_{\mathcal{I}} \rangle |\cdot  \Pi_{\langle a_{\mathcal{I}}\rangle }^{\gamma|_{\langle a_{\mathcal{I}}\rangle }}\\
		&=\frac{1}{d^n}\sum_{\mathcal{I} \subset [N]} (-1)^{|\mathcal{I}|+1} \sum_{b \in \langle a_{\mathcal{I}}\rangle } \omega^{\gamma(b)} T_b\\
		&= \frac{1}{d^n}\sum_{b \in E} \Big (\sum_{\mathcal{I} \subset [N] \, : \, b \in \langle a_{\mathcal{I}} \rangle} (-1)^{|\mathcal{I}|+1} \Big ) \omega^{\gamma(b)} T_b. 
	\end{align}
	Thus, it suffices to show that
	\begin{equation}
		\sum_{\mathcal{I} \subset [N] \, : \, b \in \langle a_I \rangle} (-1)^{|\mathcal{I}|+1} = 1 \quad \forall b \in E.
	\end{equation}
	However, this is a consequence of the \emph{inclusion-exclusion principle} \cite{Brualdi2012}, that is 
	\begin{align}
		1 =& \delta_{b \in E} = \delta_{b \in \cup_{a \in C} \langle a \rangle} = \sum_{\mathcal{I} \subset [N]} (-1)^{|\mathcal{I}|+1} \delta_{b \in \cap_{k \in I} \langle a_k \rangle }\\
		=& \sum_{\mathcal{I} \subset [N]} (-1)^{|\mathcal{I}|+1} \delta_{b \in  \langle a_\mathcal{I} \rangle }
		= \sum_{\mathcal{I} \subset [N] \, : \, b \in \langle a_{\mathcal{I}} \rangle } (-1)^{|\mathcal{I}|+1}.
	\end{align}
	Finally, we will show that there is $ c > 0$ such that we can write $ \frac{1}{d^n}\one+ c (A_\gamma - \frac{1}{d^n}\one) $ for every $ \gamma $ as a convex combination of the operators $ \frac{|\langle a_{\mathcal{I}} \rangle  |}{d^n}\Pi_{\langle a_{\mathcal{I}}\rangle }^{\gamma_{|\langle a_{\mathcal{I}} \rangle }}  \in \SP$. 
	
	Observe that the identity \eqref{eq:Agamma-as-linear-combination} is equivalent to 
	\begin{align}\label{eq:Agamma-as-linear-combination-in-trace-zero-space}
		A_\gamma - \frac{1}{d^n} \one= \sum_{\mathcal{I} \subset [N]} (-1)^{|\mathcal{I}|+1} \cdot   \frac{|\langle a_{\mathcal{I}} \rangle |}{d^n}\cdot  \Pi_{\langle a_{\mathcal{I}}\rangle^*}^{\gamma_{|\langle a_{\mathcal{I}} \rangle }} 
	\end{align}
	with 
	\begin{equation}
		\quad \Pi_{\langle a\rangle^* }^{r} := \frac{1}{|\langle a \rangle |} \sum_{b \in \langle a\rangle \setminus \{0\}} \omega^{r(b)} T_b \in \Herm_0(\mathcal{H}), \; r: \langle a \rangle \to \Z_d.
	\end{equation}
	Moreover, due to Lemma~\ref{lemma:SumsOfStabilizer CodeProjectors}, we have 
	\begin{align}
		\one = \sum_{r} \Pi_{\langle a \rangle}^r  \quad  \Longrightarrow \quad 0 = \sum_{r} \Pi_{\langle a \rangle^*}^r 
	\end{align}
	for all $a \in E, \; \frac{|\langle a \rangle|}{d^n}\Pi_{\langle a \rangle}^r \in \SP  $,
	where $ r $ ranges over all noncontextual value assignments $ r : \langle a \rangle \to \Z_d $. This implies that 
	\begin{align}
		(-1)  \cdot  \Pi_{\langle a_{\mathcal{I}}\rangle^*}^{\gamma|_{\langle a_{\mathcal{I}} \rangle }}   = \sum_{r \neq \gamma|_{\langle a \rangle}} \Pi_{\langle a_{\mathcal{I}}\rangle^*}^{r} . 
	\end{align}
	As a consequence, every summand in the right hand side of eq.~\eqref{eq:Agamma-as-linear-combination-in-trace-zero-space} can be written as a conic combination of elements $ \Pi_{\langle a \rangle^*}^r $. By properly properly rescaling, we can find $ c >0 $ such that 
	\begin{equation}
		c(A_\gamma - \frac{1}{d^n} \one) = \sum_{a} \alpha_a \Pi_{\langle a \rangle^*}^{r_a} \quad \text{with} \quad \alpha_a \ge 0, \; \sum_{a} \alpha_a = 1,
	\end{equation} 
	which is equivalent to 
	\begin{align}
		\frac{1}{d^n} \one + c(A_\gamma - \frac{1}{d^n} \one) = \sum_{a} \alpha_a (\Pi_{\langle a \rangle^*}^{r_a} + \frac{1}{d^n} \one) =  \sum_{a} \alpha_a \Pi_{\langle a \rangle}^{r_a} \in \SP.
	\end{align}
	This proves that $ c \cdot \vartriangle_{\Herm_1(\mathcal{H})} \subset \SP $ for some $ c > 0 $, which remained to be shown. 
	$ \Box $
	
\end{document}